\documentclass[twocolumn]{aastex62}

\usepackage{float}
\usepackage{rotate}
\usepackage{hyperref}

\newcommand\SPSB[2]{^{#1}_{#2}}

\begin{document}

\title{Prevalence of Complex Organic Molecules in Starless and Prestellar Cores within the Taurus Molecular Cloud}

\correspondingauthor{Samantha Scibelli}
\email{sscibelli@email.arizona.edu}

\author[0000-0002-9485-4394]{Samantha Scibelli}
\affil{Steward Observatory \\ University of Arizona \\
Tucson, AZ 85721}

\author{Yancy Shirley}
\affil{Steward Observatory \\ University of Arizona \\
Tucson, AZ 85721}

\begin{abstract}

The detection of complex organic molecules (COMs) toward dense, collapsing prestellar cores has sparked interest in the fields of astrochemistry and astrobiology, yet the mechanisms for COM formation are still debated. It was originally believed that COMs initially form in ices which are then irradiated by UV radiation from the surrounding interstellar radiation field as well as forming protostars and subsequently photodesorbed into the gas-phase. However, starless and prestellar cores do not have internal protostars to heat-up and sublimate the ices. Alternative models using chemical energy have been developed to explain the desorption of COMs, yet in order to test these models robust measurements of COM abundances are needed toward representative samples of cores. We've conducted a large-sample survey of 31 starless and prestellar cores in the Taurus Molecular Cloud, detecting methanol (CH$_3$OH) in 100$\%$ of the cores targeted and acetaldehyde (CH$_3$CHO) in 70$\%$. At least two transition lines of each molecule were measured, allowing us to place tight constraints on excitation temperature, column density and abundance. Additional mapping of methanol revealed extended emission, detected down to A$_\mathrm{V}$ as low as $\sim$ 3 mag. We find complex organic molecules are detectable in the gas-phase and are being formed early, at least hundreds of thousands of years prior to star and planet formation. The precursor molecule, CH$_3$OH, may be chemically linked to the more complex CH$_3$CHO, however higher spatial resolution maps are needed to further test chemical models.

\end{abstract}

\keywords{Astrochemistry - ISM; Star Formation - Prestellar Cores; Complex Molecules}

\section{Introduction} \label{sec:intro}

Understanding the origin, production, and distribution of complex organic molecules (COMs) and prebiotic molecules is crucial for answering astrobiologcal questions about the origins of life. Amino acids and COMs important for the formation of life are found in laboratory studies of the energetic processing of interstellar ice analogues (i.e.
\citealt{1995ApJ...454..327B}, 
\citealt{2001PNAS...98..815D}, \citealt{2002Natur.416..401B},
\citealt{1988Icar...76..225A},
\citealt{2010ApJ...718..832O},
\citealt{2011AsBio..11..847D},
\citealt{2013AsBio..13..948M},
\citealt{2017ApJ...842...52F},  \citealt{2018ApJ...865...41M},
\citealt{2018NatCo...9.5276N}, \citealt{2019MNRAS.484L.119D}).  It is beyond the current capabilities of existing observatories to remotely study this predicted complexity in interstellar ices. However, observations of the gas-phase emission spectra of COMs provide a probe of the initial primitive stages of this important chemistry.
A primary goal of the study of COMs is to understand where and how prebiotic molecules are formed in the interstellar medium, prior to potential delivery to a planetary surface.

Before a low-mass ($M$ $\leq$ few M$_\odot$) star is formed, it is conceived inside a dense clump of gas and dust known as a starless core.
Dense starless cores and gravitationally bound prestellar cores are ideal regions to study the initial stages of chemistry prior to protostar and planet formation due to their simplicity: shallow temperature gradients, absence of an internal heat source, and absence of strong shocks or outflows (\,\citealt{1989ApJS...71...89B}, \citealt{1994MNRAS.268..276W},
\,\citealt{2001ApJ...557..193E}, \citealt{2007ARA&A..45..339B},  \citealt{2014prpl.conf...27A}). 
Studies of COMs directed toward starless cores are unique in that they probe one of the earliest phases in which COMs are observed in the interstellar medium. In a few cores COMs have been detected in the gas-phase from deep observations in the 3mm band (e.g., \citealt{2012A&A...541L..12B}, \citealt{2016ApJ...830L...6J}). We still do not know just how prevalent COMs are in this phase because observations have been limited to a few well-studied cores.

The formation of COMs at the cold ($\sim$10 K) temperatures found in starless and prestellar cores is not well understood. COMs were originally believed to have been formed within icy dust grain mantles, where they would remain frozen through the prestellar phase, constructed at slow rates by UV radiation from nearby stars, the interstellar radiation field, and cosmic ray impacts (e.g., \citealt{2002ApJ...571L.173W}, \citealt{2017MNRAS.467.2552C}). COMs would then be released into the gas-phase when heating from the forming protostar would sublimate the ice mantles driving a rich, warm gas-phase chemistry (\citealt{2008ApJ...674..984A}, \citealt{2009ARA&A..47..427H}).  This grain-surface formation theory proved a better match to observations of COMs during protostellar phases, as pure gas-phase chemistry had been shown to lead to COM abundances several orders of magnitude lower than those observed \,\citep{1992IAUS..150..317C}. Slow gas-phase rates of some key reactions have been confirmed by laboratory studies. For instance, a 3\% yield of CH$_3$OH through the gas-phase dissociative recombination reaction of CH$_3$OH$_2^+$ is too inefficient to create sufficient amounts of interstellar gas-phase methanol \,\citep{2006FaDi..133..177G}.

Since there is no protostar yet during the starless core phase, a major problem in explaining the observed gas-phase COM abundances lies in understanding how COMs or their precursors can be desorbed at cold temperatures.
One possible solution is that the some COMs form in the gas-phase from reactions with precursor radical molecules (i.e. HCO, CH$_3$O) which may themselves form in the grain ice mantles and are subsequently desorbed by the chemical energy released in their formation reactions (a process called reactive desorption).  These radicals have been observed in the gas-phase toward prestellar cores \citep{2016A&A...587A.130B}. New models support the idea that precursor molecules of COMs first form on icy surfaces of interstellar grains, and then get ejected into the gas via reactive desorption (\citealt{2013ApJ...769...34V}, \,\citealt{2015MNRAS.449L..16B}, \,\citealt{2017ApJ...842...33V}). Unlike other processes, chemical desorption links the solid and gas-phase without immediate interaction with any external agents such as photons, electrons, or other energetic particles; and in this process the newly formed molecule possesses an energy surplus that allows it to evaporate\,\citep{2016A&A...585A..24M}. This process can be efficient in the cold, UV-shielded environments of prestellar cores. 
Recent experimental laboratory studies found that reactive desorption can in fact occur in the conditions found in starless core environments (\citealt{2018ApJ...853..102C}, \citealt{2018NatAs...2..228O}). Observations, like those presented in this paper, that place constraints on COM abundances are crucial to test chemical desorption models (i.e., \,\citealt{2017ApJ...842...33V}). 

Studies to date have all pointed to only a few ($<$ 10) well-known dense starless cores that may not necessarily be representative of average populations (i.e., L1544 is one of the densest, most evolved starless cores known). The lack of COM abundance measurements in a larger sample of cold cores has prevented testing of COM formation scenarios. In fact, only one prestellar core, L1544, has been thoroughly tested against chemical desorption models (\,\citealt{2016ApJ...830L...6J}, \,\citealt{2017ApJ...842...33V}). 
A study of a complete sample of starless cores within a molecular cloud is needed to constrain the question of how prevalent COMs are and what range of abundances are observed.

\begin{figure*}[!t]
\includegraphics[width=180mm]{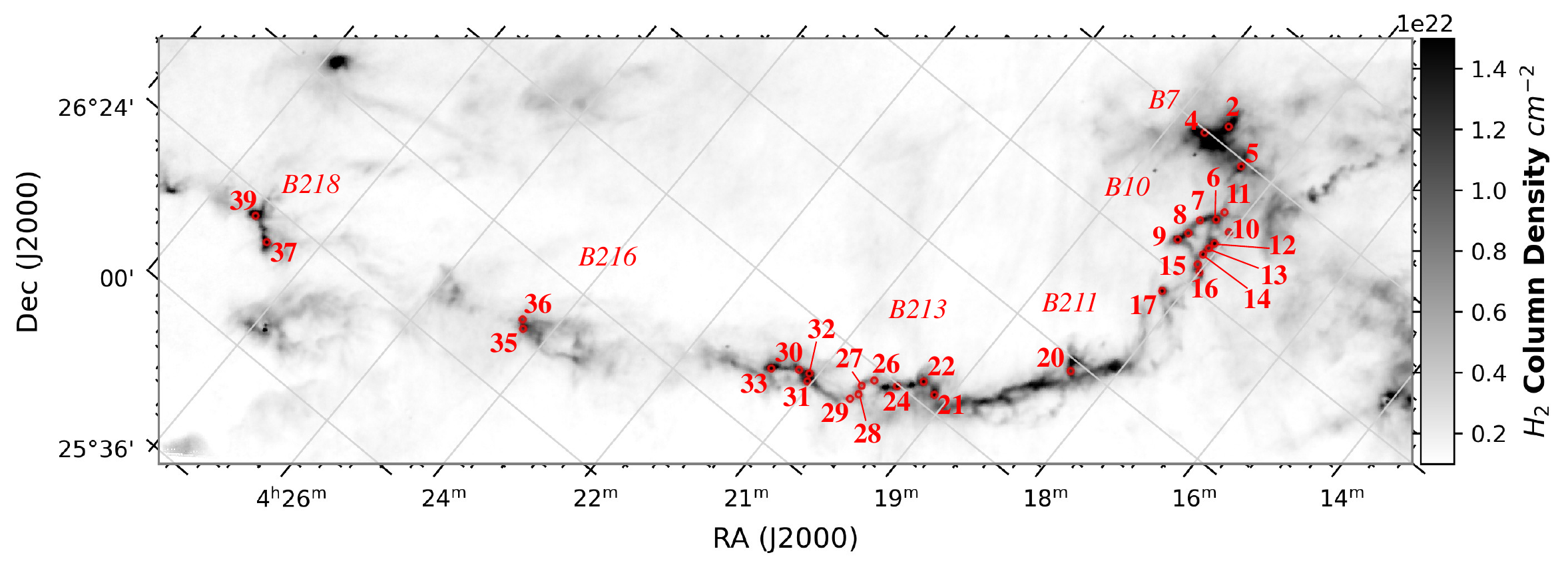} \\
\caption{ \label{coreloc} $Herschel$-derived H$_2$ column density map, presented in \citealt{2016MNRAS.459..342M}, of the B7, B10, B211, B213, B216, and B218 regions of the Taurus Molecular Cloud. The numbered circles in red represent the 31 cores targeted in this study. The circles have a diameter of 62.3$^{\prime\prime}$ (our CH$_3$OH beam size). Source selection was based on analysis of the ammonia NH$_3$ (1,1) intensity map as described in \,\citealt{2015ApJ...805..185S}. }
\end{figure*}

In this paper, we present observations of COMs in the gas-phase toward the complete ammonia identified sample of 31 starless and prestellar cores within the L1495-B218 filaments of the Taurus star forming region (Figure \,\ref{coreloc}).
We first targeted methanol, CH$_3$OH because it is one of the simplest and most abundant complex organic molecules \,\citep{2006A&A...455..577T}. We then searched for the more complex molecule acetaldehyde (CH$_3$CHO). Our study is unique in that it has targeted a large sample of starless and prestellar cores, spanning a wide range of dynamical and chemical evolutionary stages, localized within a common region within a single cloud. A survey within one molecular cloud eliminates potential chemical differences found from comparing cores from different clouds. The cores in this survey all have similar environmental conditions that warrant a more robust comparison than heterogeneous surveys. 

In section \,\ref{sec:obs} we describe the Arizona Radio Observatory (ARO) data along with our reduction techniques. We explain the source selection in section \,\ref{sourceselect}. Within section \,\ref{sec:results} we discuss our observational results, calculate column densities as well as abundances for each molecular species, and analyze the chemical and evolutionary trends. In section \,\ref{sec:discuss} we discuss the connection between the widespread precursor COM, methanol, and acetaldehyde.

\section{Observations and Data Reduction}\label{sec:obs}

\begin{figure*}
\centering
\begin{center}
\includegraphics[scale=0.5]{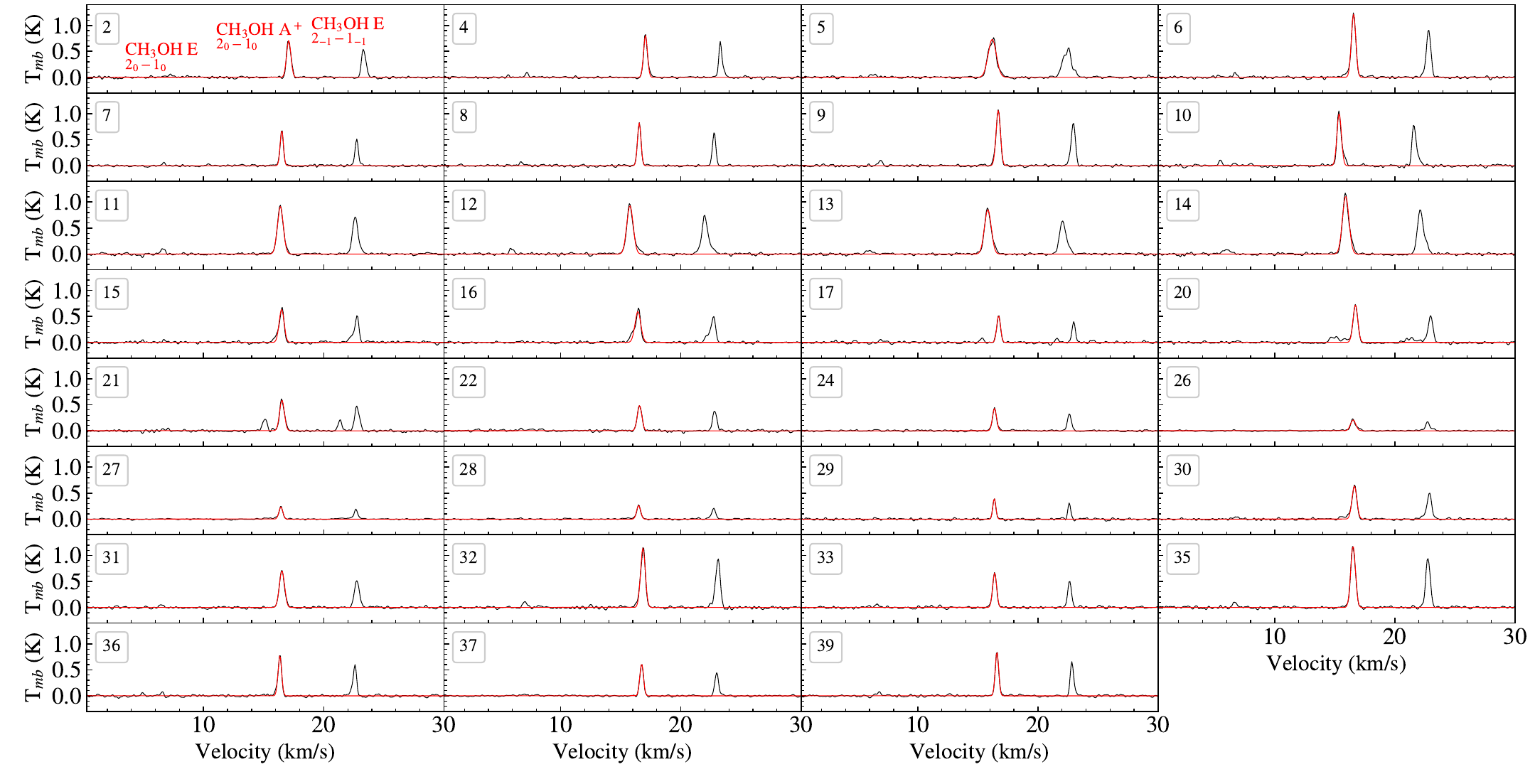}
\end{center}
\caption{\label{taurus_met_spec}Spectra of methanol in the 31 prestellar cores in Taurus. Overplotted in red is the fit to the brightest CH$_3$OH A$^+$ line. Numbers in the top left correspond to the regions labeled in Figure \,\ref{coreloc}. Note that the plotted velocity correspond to v$_{LSR}$ of the weakest methanol line.
}
\end{figure*}

\begin{figure*}
\centering
\begin{center}
\includegraphics[scale=0.5]{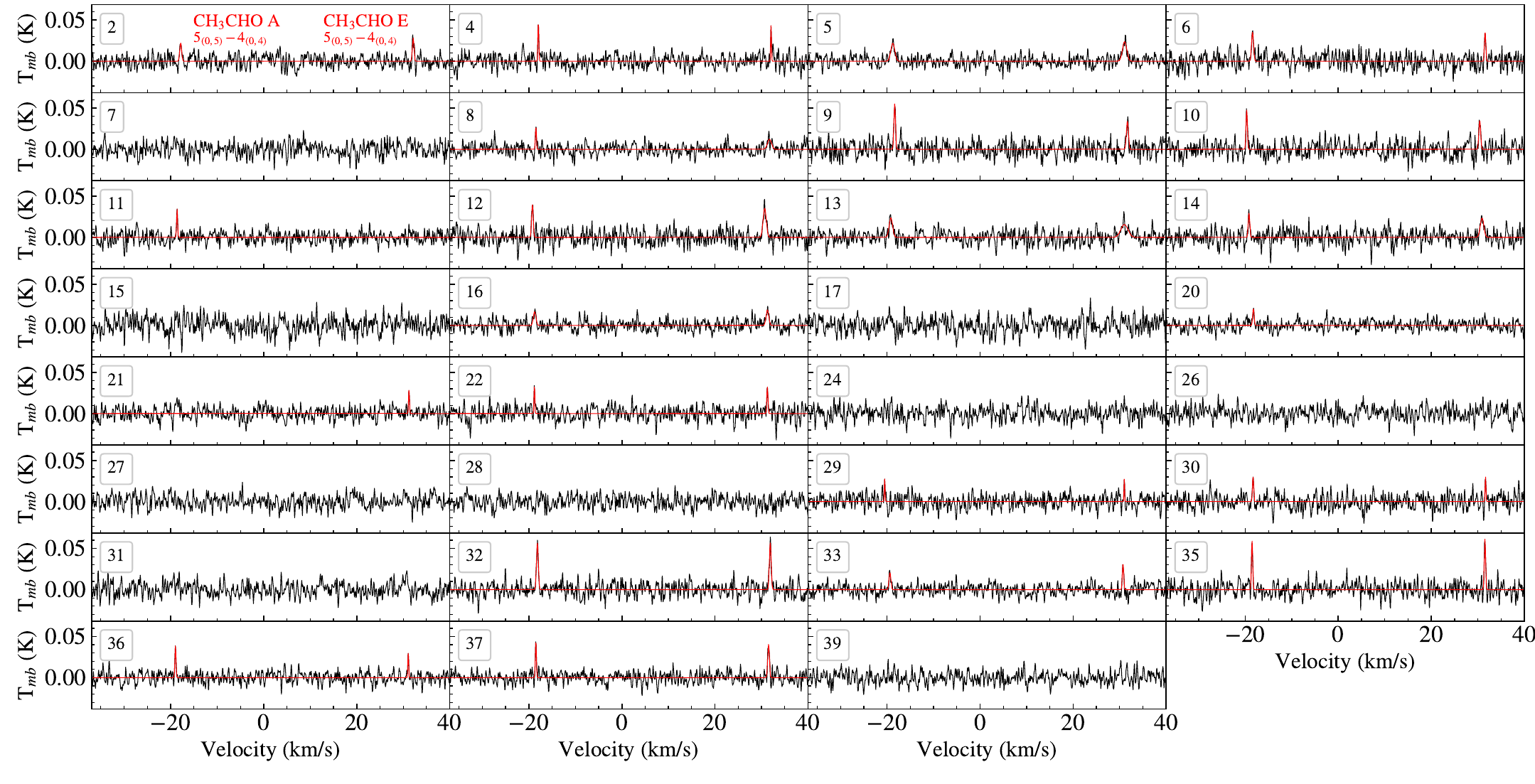}
\end{center}
\caption{\label{taurus_acet_spec}Spectra of the 96 GHz transitions of CH$_3$CHO in all 31 prestellar cores in Taurus (including non-detections). Numbers in the top left correspond to the regions labeled in Figure \,\ref{coreloc}. Overplotted in red are fits to the detected lines. Spectra with no red fits are considered non-detections. Note that the plotted velocity is the v$_{LSR}$ of the average of the A and E line frequencies. } 
\end{figure*}

\begin{figure}[tbh]
\includegraphics[scale=0.46]{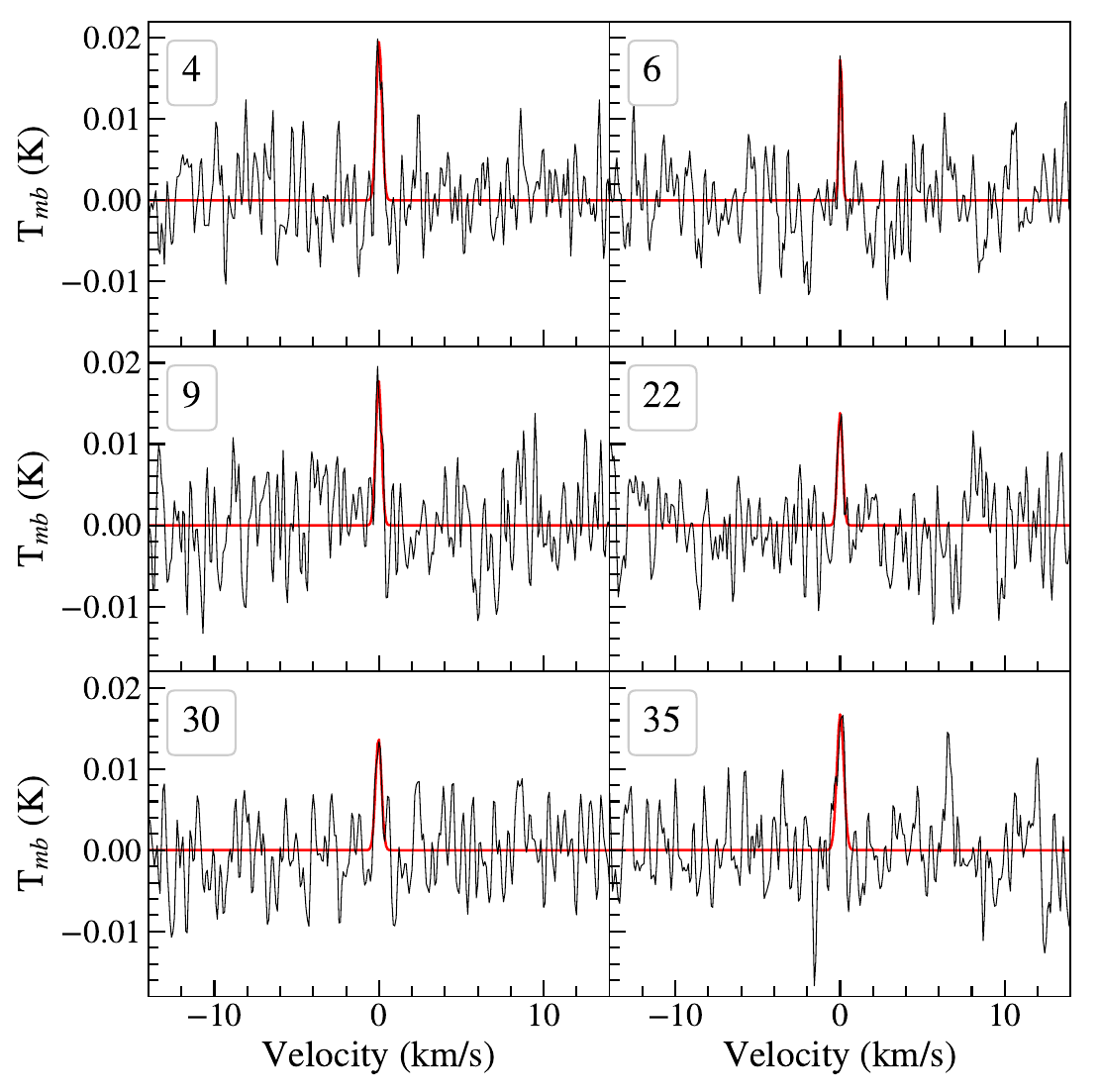}
\caption{\label{ch3choc_spectra} Spectra of the 84 GHz line of CH$_3$CHO and corresponding Gaussian fit in red for the 6 detected sources. The expected frequency of this transition was shifted to a v$_{LSR}$ of 0 km s$^{-1}$ for visual comparison.
Numbers in top left correspond to the cores labeled in Figure \,\ref{coreloc}.}
\end{figure}

Molecular line observations for 31 cores within the L1495-B218 filament in the Taurus Molecular Cloud were taken with the ARO 12m telescope on Kitt Peak. From previous ammonia survey results, we know these cores have kinetic temperatures of 8-11 K, thermally-dominated velocity dispersions of 0.08-0.24 km s$^{-1}$, and dust masses ranging from 0.05 to 9.5 M$_\odot$ (Table 1 \& 2 of \,\citealt{2015ApJ...805..185S}). Single pointing observations were carried out from October 2017 to March 2018 in 50 shifts totaling around 411 hours. Three transitions of 2$_k$-1$_k$ lines of methanol were targeted; CH$_3$OH E 2$_{-1}$-1$_{-1}$ ($E_u/k$ = 12.5 K) centered at 96.739 GHz, CH$_3$OH A$^+$ 2$_{0}$-1$_0$ ($E_u/k$ = 6.9 K) centered at 96.741 GHz, and CH$_3$OH E 2$_{0}$-1$_0$ ($E_u/k$ = 20.1 K) centered at 96.744 GHz. Additionally, three transitions of acetaldehyde (5$_{(0,5)}$-4$_{(0,4)}$ E and A as well as the 2$_{(1,2)}$-1$_{(0,1)}$ lines) were targeted (parameters listed in Table\,\ref{LineParam}). Given a FWHM of 62.3$^{\prime\prime}$, the beam radius for our methanol observations was 31.15$^{\prime\prime}$, or 0.02 pc, at a distance of 135 pc (\citealt{2014ApJ...786...29S}).
Each scan was 5 minutes using absolute position switching (APS) between the source and the off position every 30 seconds. Observations of each source roughly took $\sim 1$ hour, $\sim 3$ hours, and $\sim 6$ hours for the 96 GHz lines of CH$_3$OH, CH$_3$CHO, and the 84 GHz line of CH$_3$CHO, respectively. We pointed at the NH$_3$ peak position as tabulated in \citealt{2015ApJ...805..185S}. Pointing was checked every hour on a nearby quasar or planet. The system noise temperature was $\sim$ 150 K. The MAC (Millimeter Auto Correlator) instrument was used as the back-end, with a bandwidth of 150 MHz and 24.4 kHz resolution with Hanning smoothing. The frequency resolution corresponds to a velocity resolution of $\sim$0.08 km s$^{-1}$ at 96 GHz, which is narrower than the expected line width of these cores ($\sim$0.3 km s$^{-1}$). These were dual polarization observations, providing two independent, simultaneous observations used to distinguish between false spectral features and real lines. Single pointing spectra are plotted in Figures \,\ref{taurus_met_spec}, \,\ref{taurus_acet_spec}, and \,\ref{ch3choc_spectra}.

\begin{deluxetable}{ccccc}
\tablecaption{Line Parameters \label{LineParam}}
\tablewidth{0pt}
\tablehead{ \colhead{Molecule}& \colhead{Transition}& \colhead{Freq. } &\colhead{E$_{up}$} &\colhead{A$_{ij}$} \\ \colhead{ }& \colhead{}& \colhead{(GHz)} &\colhead{(K)} &\colhead{(s$^{-1}$)}}
\startdata
\hline
CH$_3$OH   &  2$_{-1,2}$ - 1$_{-1,1}$  E& 96.739363 & 12.542 & 2.557E-6 \\
     &  2$_{0,2}$ - 1$_{0,1}$ A$^+$ & 96.741377 & 6.965 &  3.408E-6  \\
   &  2$_{0,2}$ - 1$_{0,1}$ E & 96.744549  & 20.090  & 3.408E-6  \\
 CH$_3$CHO &  5$_{0,5}$ - 4$_{0,4}$ E  & 95.947439  & 13.935  & 2.955E-5   \\
    & 5$_{0,5}$ - 4$_{0,4}$ A  &  95.963465  & 13.838  & 2.954E-5 \\
   &    2$_{1,2}$ - 1$_{0,1}$ A$^{++}$ & 84.219750  &  	4.967 &  2.383E-6  
\enddata
\tablecomments{Taken from SLAIM\,\footnote{http://www.cv.nrao.edu/php/splat/}.}
\end{deluxetable}

Data reduction was performed using the CLASS program of the GILDAS package\,\footnote{http://iram.fr/IRAMFR/GILDAS/}. Peak line fluxes, intensities, velocities and linewidths have been derived by Gaussian fits to the line profiles. Two separate beam efficiencies were determined from 34 planet (Jupiter, Mars and Venus) observations taken over the course of our observation run. The median efficiency percentage for each planet was calculated, along with estimated errors, for each polarization (called MAC11 and MAC12). The median beam efficiency value for the combination of the three planets was then applied to our observations in each separate channel. These channels were then summed together, giving us the main beam temperature scale; $T_{mb}$ = $T_A^*$/$\eta$, where $\eta_{MAC11}$ = 0.861$\pm$0.006 and $\eta_{MAC12}$ = 0.838$\pm$0.006. Additionally, a factor of 1.14 needed to be multiplied to the final main beam temperature due to a systematic calibration error present in the software of the MAC, discovered after our APS observations were completed. The MAC antenna temperature was found to be 14$\%$ lower than the new more trusted AROWS back-end. This result only became available in April 2018 after APS observations were complete.

\begin{figure*}
\centering
\begin{center}$
\begin{array}{l}
\includegraphics[width=180mm]{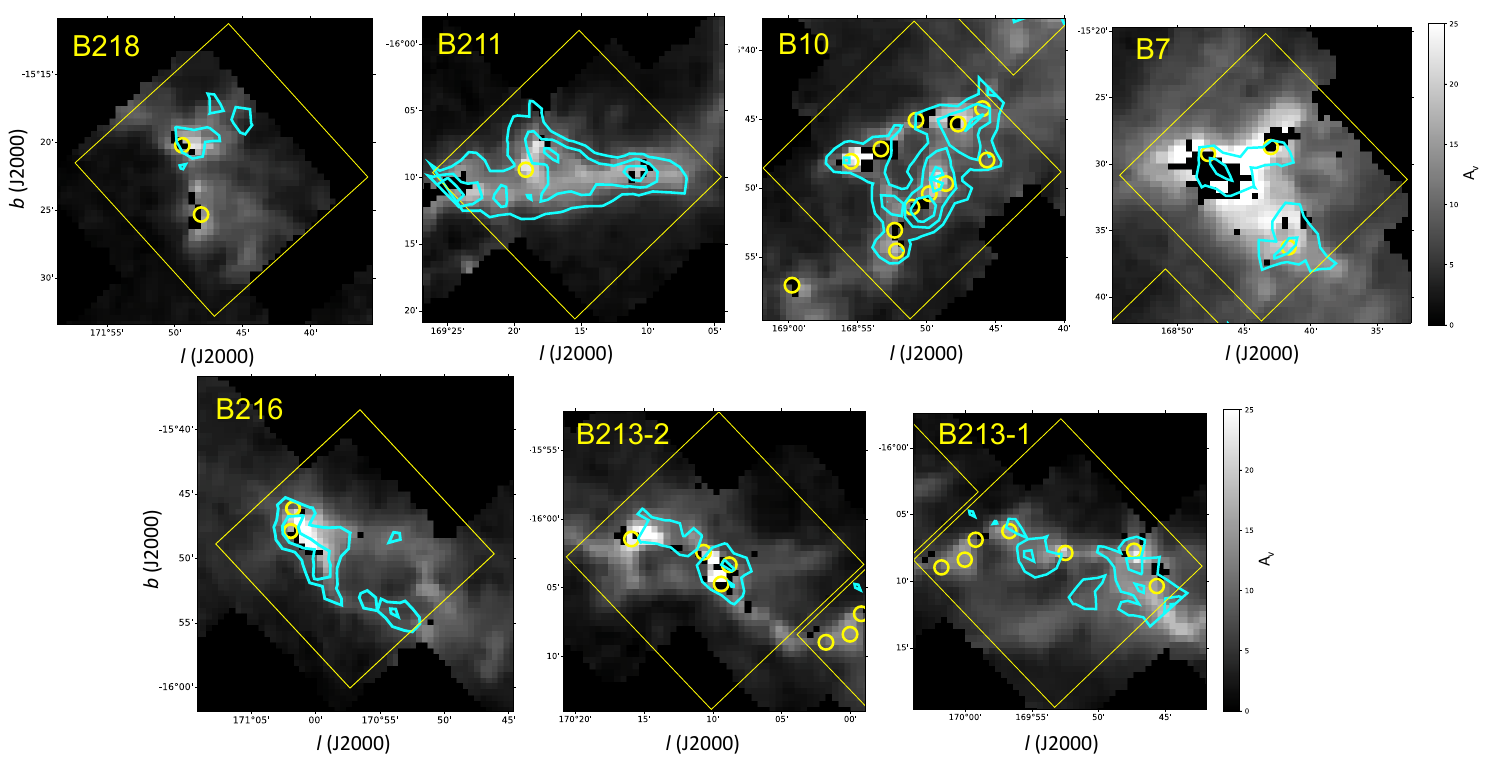}
\end{array}$
\end{center}
\caption{ \label{extmap} Methanol emission (cyan contours) overlaid on extinction maps \citep{2010ApJ...725.1327S} of the B7/B10/B211/B213/B216/B218 regions of Taurus (axes in galactic coordinates $l$ and $b$). The yellow boxes correspond to the 15$^{\prime}$ $\times$ 15$^{\prime}$ size of our OTF maps and the yellow circles represent the cores targeted in this study. The cyan contours within the boxes are plotted in steps of 0.2 K km s$^{-1}$ ($\sim$2$\sigma$ detection) created from the integrated intensity maps at a resolution of 81$^{\prime\prime}$.  }
\end{figure*}

Before the new AROWS spectrometer was commissioned in April 2018, On-The-Fly (OTF) mapping was not possible with the MAC because data rates were too high (i.e., data could not be taken and stored simultaneously). We began the mapping of CH$_3$OH emission in May 2018. Seven separate maps were observed towards regions where most of the cores are located. We tuned to 96.741375 GHz in the lower sideband choosing the 19.5 kHz mode on AROWS using only the central 1024 channels. We performed OTF mapping in 15$^{\prime}$ x 15$^{\prime}$ regions (see Figure \,\ref{extmap}), with each row spaced at 22$^{\prime\prime}$. The scan rate was at 15$^{\prime\prime}$ per second with OFF integration time at 36 seconds and calibration integration time at 5 seconds. We made multiple maps in RA and DEC directions, later baselining and combining the maps within the CLASS software. For intensity mapping, data were processed with a pipeline script in CLASS written by W. Peters (see \citealt{2011ApJS..196...18B}) which created 3D spectral cubes with a new convolved beam size of 81.17$^{\prime\prime}$. The program miriad\,\footnote{https://bima.astro.umd.edu/miriad/} was then used to create integrated intensity (moment 0) maps (Figure \,\ref{extmap}). The re-sampling of OTF maps at lower resolution is needed so as to not miss information in the image field and to improve signal-to-noise.
We present only single pointing measurements for CH$_3$CHO because it was not bright enough to map, i.e., it would have taken $\sim$2,000 hours to map the same $15' \times 15'$ regions as we did for methanol or $\sim$5,000 hours to map each core in $2' \times 2'$ regions to get down to 4 mK $rms$.

\section{Source Selection} \label{sourceselect}

Sources were chosen from cores defined from the Cardiff Source-finding AlgoRithm (CSAR), as described in \citealt{2013MNRAS.432.1424K}, performed on an ammonia NH$_3$ (1,1) intensity map, as described by \citealt{2015ApJ...805..185S}. From the total 39 listed cores (leaves in the CSAR output) we discarded 4 sources that were observed to have protostars, 3 sources where the 12m beams overlapped, and 1 source (L1495A-S) that had been previously studied in \citealt{2016A&A...587A.130B}.

Calculations of column density and excitation temperature for the COMs detected requires average volume density measurements for each core. Using our single pointing beam size of 62.3$^{\prime\prime}$ (8410 AU at 135 pc; \citealt{2014ApJ...786...29S}), we calculated the average volume density within each of the 31 cores using the line-of-sight distance of 135 pc and a median H$_2$ column density ($N_{H_{2}}$) from $Herschel$ column density maps of Taurus, which have 18$^{\prime\prime}$ resolution (\,\citealt{2013A&A...550A..38P}, \citealt{2016MNRAS.459..342M}). 
The median dust temperature from the corresponding $Herschel$ temperature maps was also calculated. Within Ds9 \citep{2003ASPC..295..489J} we overlaid region files of our beam size onto the maps and recorded median H$_2$ column density ($N_{H_{2}}$) and dust temperature ($T_{dust}$) values (Table \,\ref{physparams}). Due to limited resolution, we stress that by averaging core properties over the beam we are making global measurements, since a detailed physical model of the sources from high resolution (10$^{''}$) dust data does not yet exist. Central core densities can be an order of magnitude denser than local volume density measurements, i.e., Seo12 is believed to have a central density of $\sim$10$^6$ cm$^{-3}$ \citep{2019PASJ...71...73T}, while it's beam averaged density reported in this paper is $\sim$10$^5$ cm$^{-3}$. Thus, the high uncertainties of physical source parameters (including kinematics as well as density and temperature profiles) make direct comparison between sources challenging if one attempted more advanced modeling techniques. The average ratio between our beam averaged volume density and the volume density reported in \citealt{2015ApJ...805..185S} is $<\frac{n_{Beam}}{n_{seo}}>$ = 0.81 with a standard deviation of 0.19 (median is 0.76 with a median standard deviation of 0.14). In \citealt{2015ApJ...805..185S} they assumed kinetic temperature was equal to the dust temperature, which is not true. Since SED fitting went into the calculation of the column density maps in the $Herschel$ data, we use $n_{Beam}$ in this paper.  The NH$_3$ (1,1) observations in \citealt{2015ApJ...805..185S} only effectively probe density $> 10^3$ cm$^{-3}$ \citep{2015PASP..127..299S}.

\begin{figure}[tbh]
\centering
\begin{center}$
\begin{array}{cc}
\includegraphics[scale=0.52]{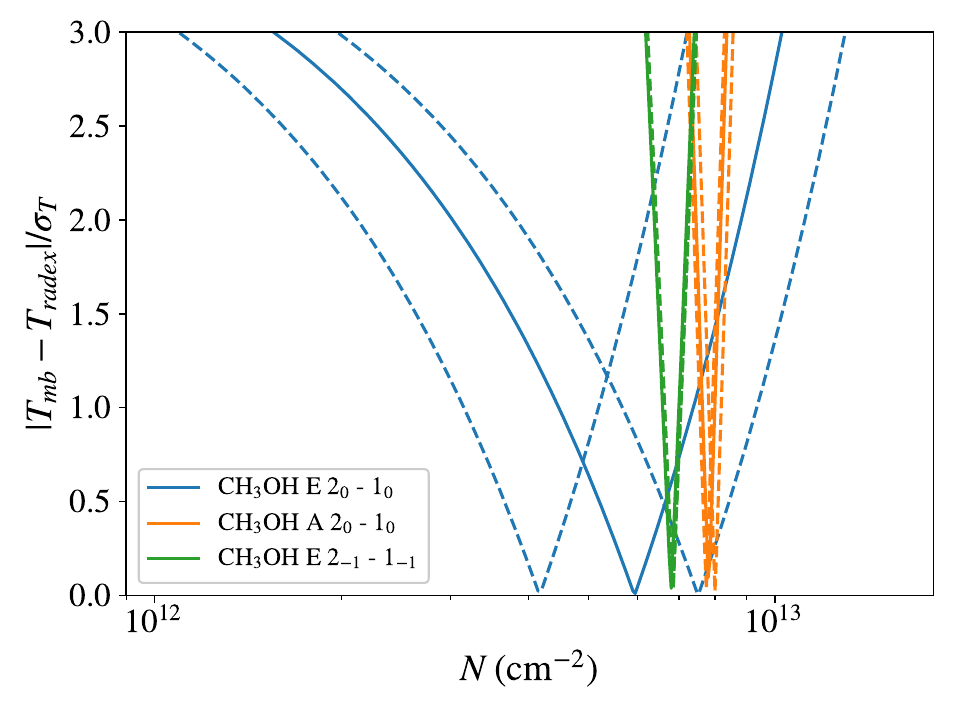} \\
\includegraphics[scale=0.53]{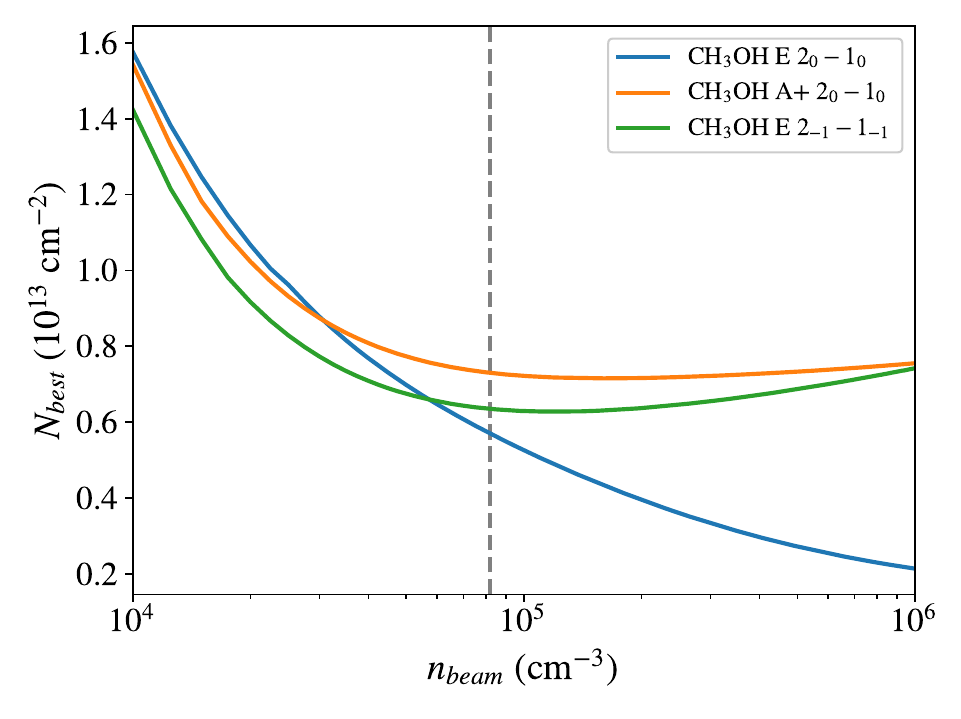} 
\end{array}$
\caption{\label{ch3ohTex} (Top panel) Difference in observed versus modeled radiation temperature divided by the $rms$ of our data ($|T_{mb}-T_{radex}|/\sigma_T$) is plotted against column density from RADEX models for Seo15.  Dashed curves represent the error as determined from grids of RADEX models run for the extrema of the errors that go into the calculation (volume density, kinetic temperature, etc.,). (Bottom panel) We plot how column density changes with varying inputs of volume density for core Seo15. For the two bright CH$_3$OH transitions (orange and green lines), a span of two orders of magnitudes variation in the volume density within the beam will result in only a factor of two variation in column density. The beam-averaged volume density for Seo15 is plotted as a grey dashed line. }
\end{center}
\end{figure}

\begin{figure*}
\centering
\begin{center}
\includegraphics[scale=0.5]{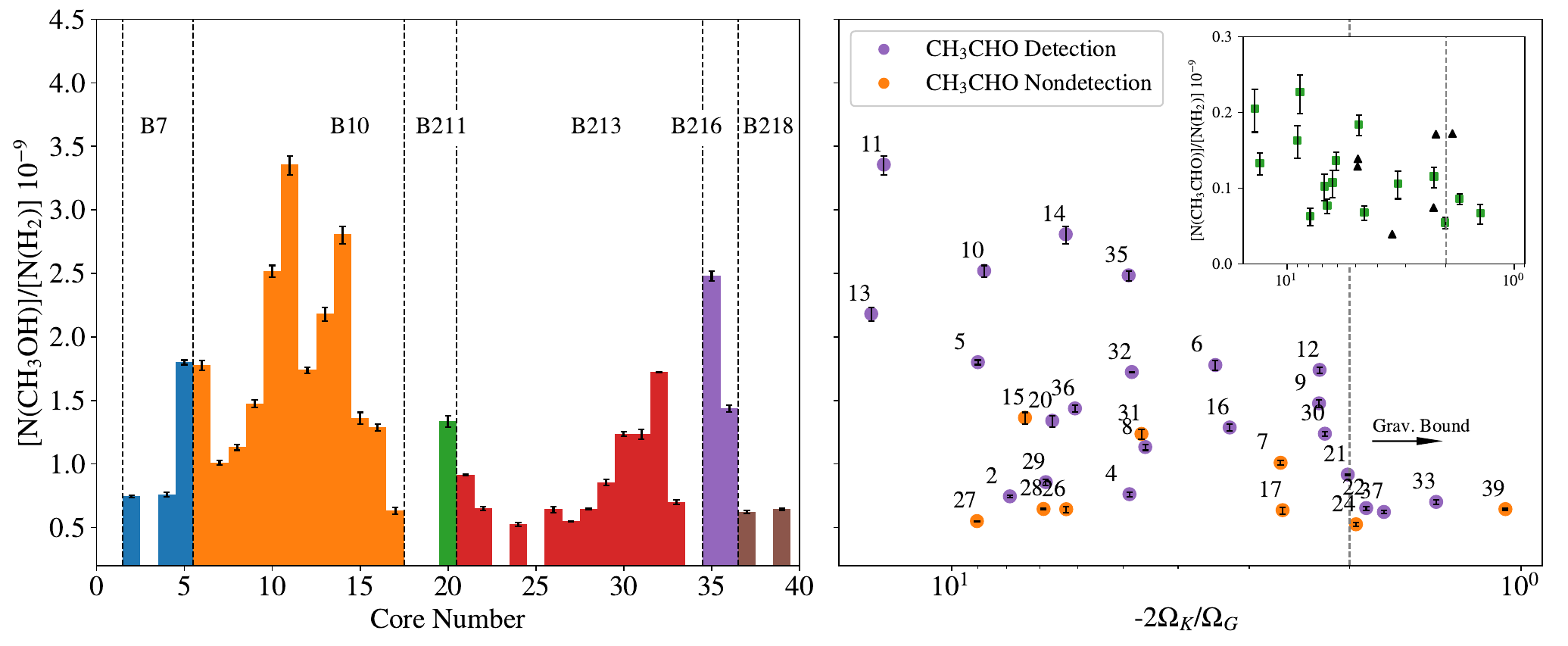} 
\caption{\label{abund_corenum} (Left panel) Plot of methanol abundance vs. core, color coded for each Barnard region that the cores are located in. The less-evolved regions (determined by lack of protostellar sources), i.e., B10, B211 and B216, have higher methanol abundances compared to the other more-evolved regions. (Right panel) Abundances vs. virial ratio, $\alpha =$ 2$\Omega_K/\Omega_G$ $= 5\sigma^2 R/GM$. In orange we denote which cores CH$_3$CHO was not detected. Within the plot insert black triangles represent the 6 cores for which two transitions of CH$_3$CHO were detected and whose median $T_{ex}$ value was used to calculate column density for the remaining cores. Error bars for these 6 cores have been removed due to large dispersion (see Figure \,\ref{ch3chocTex}). Note: the x-axis increases towards the left. }
\end{center}
\end{figure*}

\begin{figure}[tbh]
\begin{center}$
\begin{array}{c}
\includegraphics[scale=0.38]{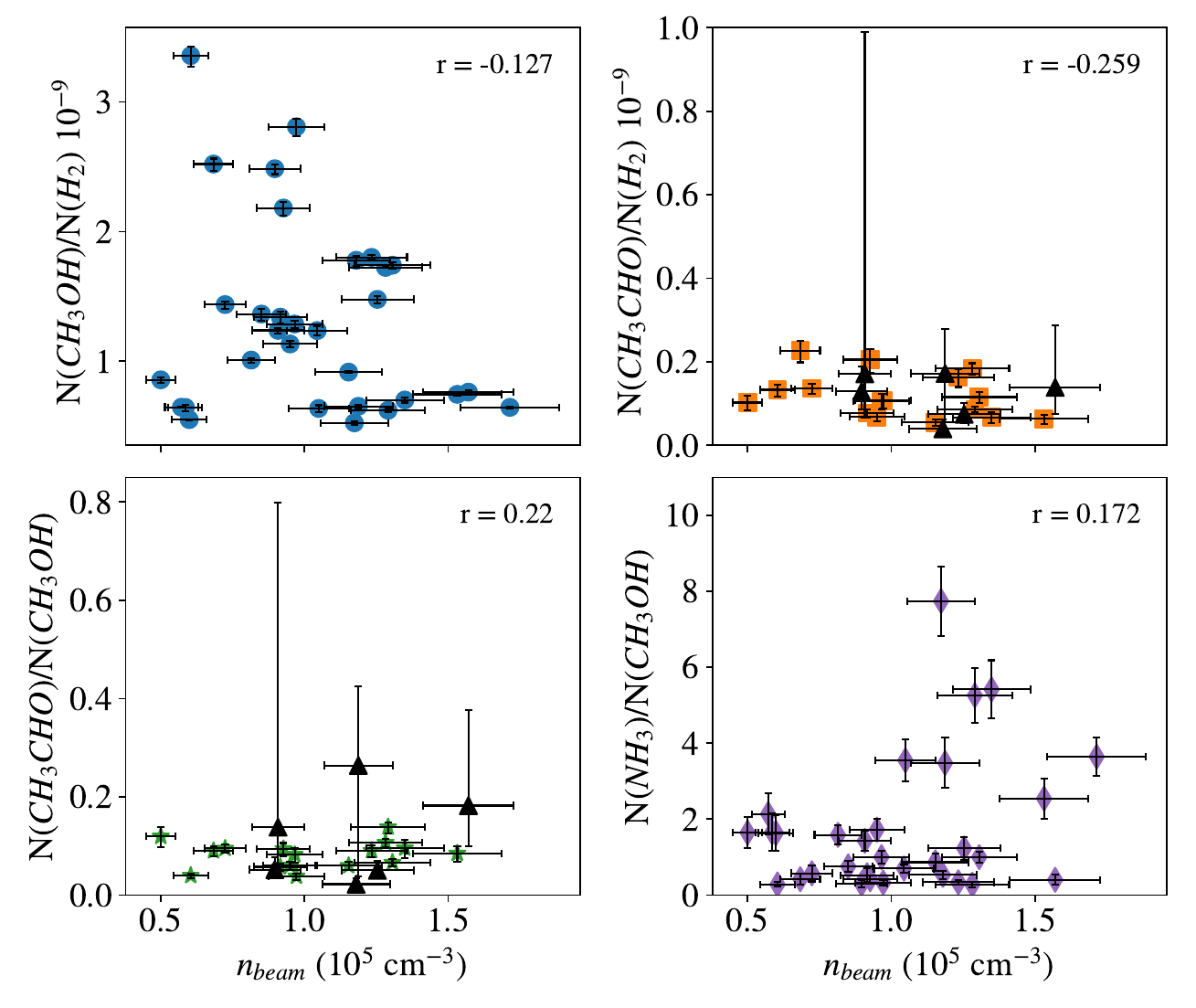}
\end{array}$
\end{center}
\caption{\label{textvsnbeam} Plots of abundance and abundance ratios versus volume density, $n_{beam}$. The top left panel plots  CH$_3$OH abundance wrt H$_2$ as blue circles, the top right plots CH$_3$CHO wrt H$_2$ as orange squares, the bottom left plots  CH$_3$CHO wrt  CH$_3$OH as green stars and the bottom right plots NH$_3$ wrt CH$_3$OH as purple diamonds (NH$_3$ values from \citealt{2015ApJ...805..185S}).The black triangles symbolize the cores we detected both CH$_3$CHO transitions in, as in Figure \,\ref{abund_corenum}. The core with the largest error bar is Seo30, whose column density was poorly constrained by the CTEX method (see Figure \,\ref{ch3chocTex}). We report the Spearman rank correlation coefficient in the upper right of each panel.}
\end{figure}

\begin{figure*}
\centering 
\includegraphics[scale=0.74]{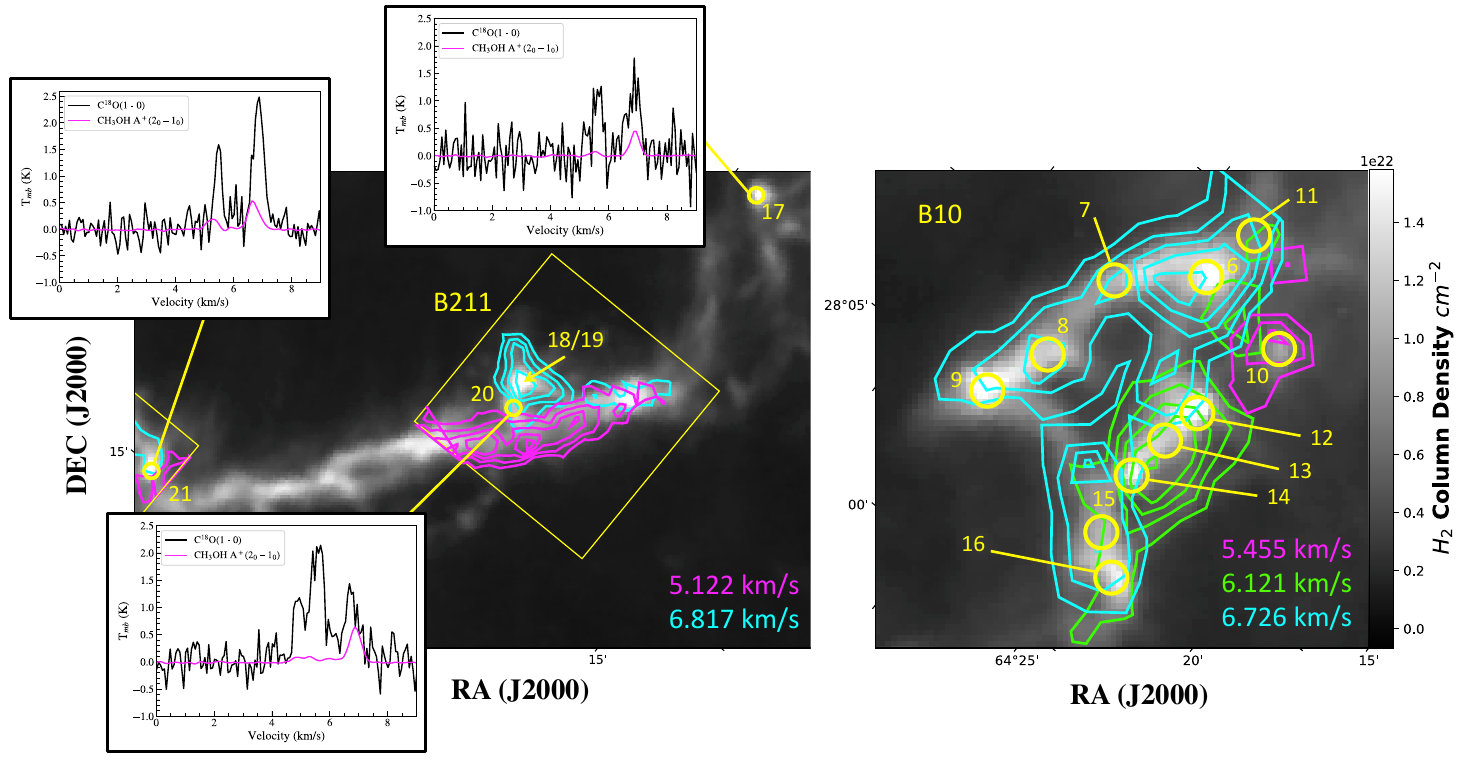}
\caption{\label{velcontours} Maps (greyscale is H$_2$ column density) of the (left) B211 and (right) B10 regions illustrating the complex velocity structure of methanol emission. On the left map three panels of spectra are shown, with C$^{18}$O (1-0) molecular observations from the IRAM 30m telescope in black \citep{2013A&A...554A..55H}. For comparison, we include the positions of the methanol peaks which shows similar velocity structure as C$^{18}$O (1-0). We shifted the bright A$^+$ to the center to show how the v$_{LSR}$'s compare, showing some misalignment perhaps due to gas motions within the filament. The yellow circles represent cores targeted in those regions. In B211 we also note with a yellow arrow where cores Seo18 and Seo19 lie from \citealt{2015ApJ...805..185S} (these weren't targeted for APS measurements due to overlapping beams). For both regions main beam temperature contours (in steps of 0.2 K starting at 0.2 K) were created in cyan at the velocity of $\sim$7 km s$^{-1}$, in lime at the velocity of $\sim$6 km s$^{-1}$ and in magenta the velocity of $\sim$5 km s$^{-1}$. 
}
\end{figure*}

\begin{figure}
\centering
\begin{center}$
\begin{array}{cc}
\includegraphics[width=85mm]{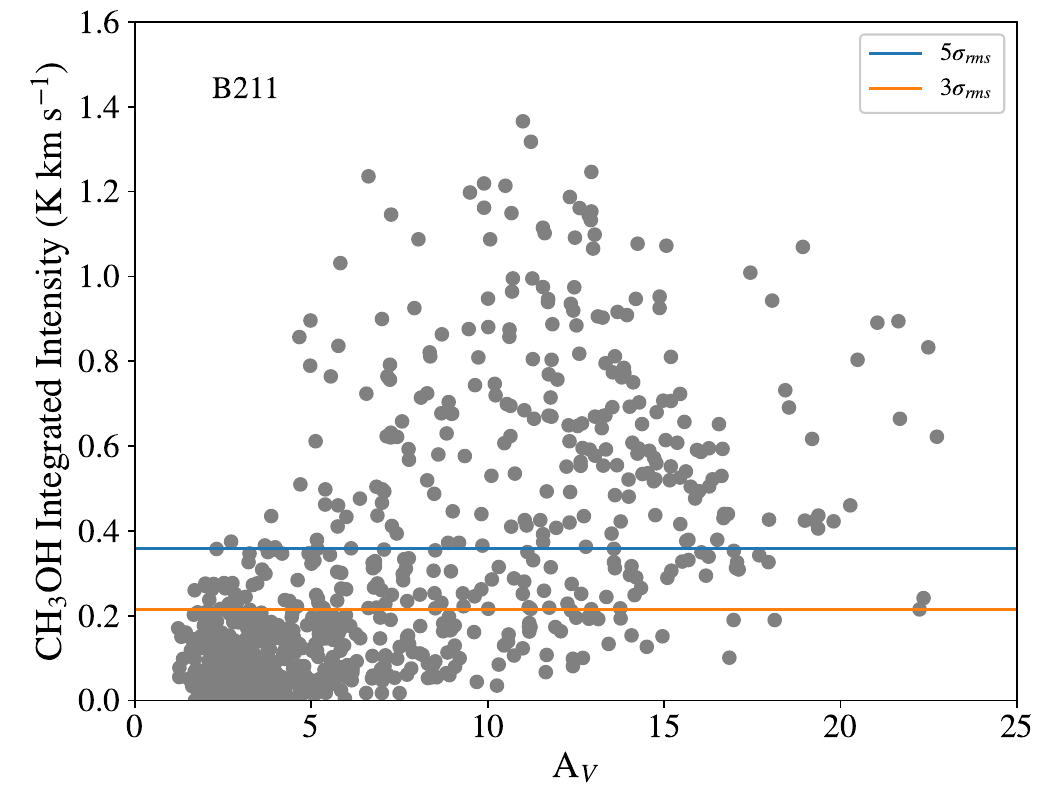} \\
\includegraphics[width=85mm]{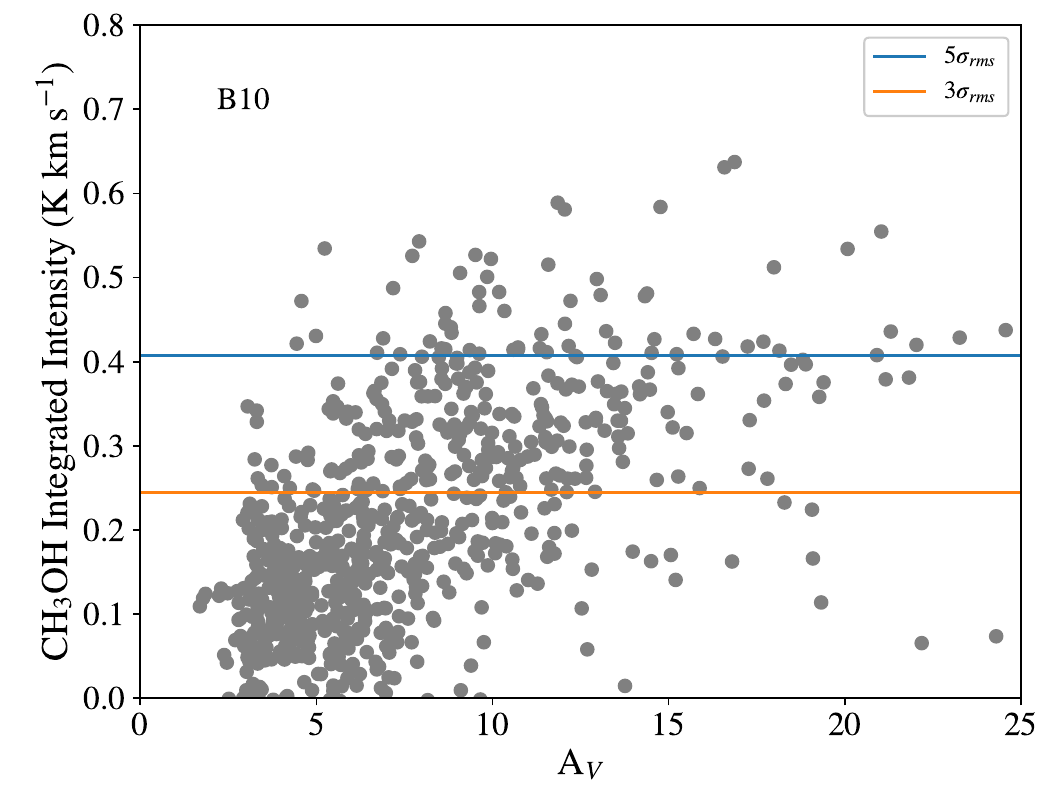}
\end{array}$
\end{center}
\caption{ \label{AVscatter} Brightest methanol peak intensity versus A$_\mathrm{V}$ for (top) B211 and (bottom) B10 regions in Taurus. B211 is the only region used to create a reliable, contiguous map, created from integrated intensities that are $\geq$5$\sigma$. } 
\end{figure}

\section{Results}\label{sec:results}

We detected CH$_3$OH in 100$\%$ (31/31) of the cores targeted and CH$_3$CHO in 70$\%$ (22/31). In the following subsections we discuss how we calculated molecular column densities for all cores and further discuss the distribution of methanol from our OTF mapping results. All values in Tables\,\ref{physparams}--\,\ref{Abundances} are from the more sensitive single pointing observations.

\subsection{CH$_3$OH Column Densities} \label{metcoldensec}

Methanol was detected in all 31 cores (Figure \,\ref{taurus_met_spec} and Table \,\ref{MethFits}). The $v_{LSR}$ of the methanol lines is consistent with that of the ammonia NH$_3$ (1,1) line and the ratio of NH$_3$ (1,1) to the brightest CH$_3$OH A$^+$ 2$_{0}$-1$_0$ transition is on average 1.02. The median \textit{rms} noise in the spectra is $\theta_{mb}\sim$15 mK. CH$_3$OH E 2$_{0}$-1$_0$ is the weakest of the three lines, due to its higher upper energy, and therefore not always detected. If this line was not detected above $>$4$\sigma$ we present upper limits (Table\,\ref{MethFits}). As a test, we integrated on core Seo26 (one of the weak detection cores) four times as long ($\sim$4 hrs vs. $\sim$1 hr) to lower the \textit{rms} to $\sim$6 mK to see if the weaker methanol line could be detected. Unfortunately, we were not able to confirm a detection and an upper limit is still reported.

The brightest cores, determined from the brightest CH$_3$OH A$^+$ 2$_{0}$-1$_0$ transition, are (from brightest to weakest) Seo6, Seo32, Seo35, Seo14 and Seo9. The two brightest CH$_3$OH lines are clearly detected in all cores (to the 12-78$\sigma$ level). The detection of more than one line of methanol shows the unambiguous presence of methanol in the cold gas within NH$_3$-detected starless cores. 

Since we detected multiple lines of CH$_3$OH with different $E_u/k$ values, we used the radiative transfer code RADEX to calculate column densities (see \citealt{2007A&A...468..627V}). RADEX calculates an excitation temperature, $T_{ex}$, a column density, $N$, and an opacity, $\tau$, for each transition separately. A grid of RADEX models was created to find the best-fit column density. The difference in our observed line peak versus the line peak RADEX calculates was minimized in order to find the best fit. In the top panel of Figure \,\ref{ch3ohTex} we show an example for core Seo15, where we plot the difference in radiation temperatures divided by the observed rms, written as $|T_{mb}-T_{radex}|$/$\sigma_T$, versus the column density $N$ for each transition line. The best-fit column density for all three methanol transitions fall within a factor of at most 1.3 of each other. 

In RADEX calculations there are three input parameters, volume density, $n_{beam}$, gas kinetic temperature, $T_{kin}$, and linewidth, $FWHM$, that were varied to get error estimates for column density. We input the average beam volume densities described in section\,\ref{sourceselect} into our RADEX calculations. The statistical volume density error was estimated to be 10$\%$ of our value, based on typical estimates of the statistical uncertainty in calculations of the column density of H$_2$ from Herschel maps (see \citealt{2013MNRAS.432.1424K}). We point out that the dust opacity assumption could easily lead to a factor of 2 to 3 in the systematic uncertainty in volume density
(\citealt{2011ApJ...728..143S}). However, we find that even if our volume density calculations are off by an order of magnitude, we would only be a factor of $\sim$2 off in column density, as determined from RADEX calculations (Figure \,\ref{ch3ohTex}, bottom panel). The statistical error for $T_{kin}$ came directly from Table 2 in \citealt{2015ApJ...805..185S}, and the statistical FWHM error came from our CLASS Gaussian fits (Table\,\ref{MethFits}). In a test case we plotted all 27 statistical error combinations (each parameter having a plus and minus error) and found that the combination of all `plus' values and the combination of all `minus' values gave the widest difference in column density in the grids. Therefore, we adopted this error combination when calculating our statistical errors for the remaining cores.

In general all lines have been minimized around a similar column density (within a factor of $\sim$1.3). However, in some cases we did not detect the third line or the third line was only an upper limit so our column density becomes uncertain (large errors). In Table\,\ref{N_tex_tau} we present this minimized, or `best-fit', column density for all three transitions separately, as well as a total column density, $N_{tot}$, which is a sum of the two brightest transitions (excluding the weakest E state).

As a consistency check, we compared our RADEX-determined column densities to the commonly used method for calculating column density from an optically thin line.  The CTEX method (Constant T$_{ex}$) assumes a constant excitation temperature when converting from the column density in the upper level of the transition to all energy levels (see details in Appendix of \citealt{2002ApJ...565..344C} and Equation 80 of \citealt{2015PASP..127..266M}).
We calculated CTEX column densities for the brightest CH$_3$OH A$^+$ 2$_{0}$-1$_0$ transition assuming the $T_{ex}$ determined from RADEX calculations. We found that RADEX-determined column densities all agree with CTEX-determined column densities, the median ratio is 1.08 with a median standard deviation of 0.03.

The total column density, $N_{tot}$ from Table\,\ref{N_tex_tau}, ranges from 0.42 -- 3.4 $\times$ 10$^{13}$ cm$^{-2}$. Excitation temperatures range from $T_{ex} =$ 6.8 -- 8.7 K, and optical depths that are consistent with optically thin
$\tau < 0.3$ (see Table \,\ref{N_tex_tau}). We found that on average the A:E column density ratio is 1.3 with a median absolute standard deviation of 0.03. This ratio agrees with the \citealt{2019arXiv190311298H} A:E ratio of 1.2 -- 1.5 observed for the starless core H-MM1 in Ophiuchus.

\subsubsection{CH$_3$OH Abundance Trends}

We found total CH$_3$OH abundances (A+E species with respect to H$_2$) for the 31 cores ranging from 0.525 -- 3.36 $\times 10^{-9}$.  Our observed abundances are comparable to published values toward other prestellar cores \citep{2006A&A...455..577T, 2014ApJ...795L...2V, 2015ApJ...802...74S, 2018arXiv180200859P}. The \citealt{2015ApJ...805..185S} paper analyzed which regions in L1495-B218 were more or less evolved by searching for the presence of protostars (Class 0, I, or II) within the regions. They find the regions B7, B213 and B218 are more evolved and regions B10, B211 and B216 are less evolved (containing only starless and prestellar cores). In the left panel of Figure \,\ref{abund_corenum} we find that the less evolved regions typically have a higher methanol abundance, i.e., a median methanol abundance (wrt to H$_2$) of 1.48 $\times$ 10$^{-9}$, versus the more evolved regions which have a median abundance 0.72 $\times$ 10$^{-9}$. This result is consistent with the picture that the methanol has `peaked' away (i.e., has a maximum abundance offset) from the center of the core in more evolved regions (as seen in L1544; \citealt{2014A&A...569A..27B}, \citealt{2018arXiv180200859P}), and we have probed the regions where methanol is depleted within a significant fraction of our beam.

We plot calculated abundances versus the virial parameter $\alpha$, which tells us if our cores are gravitationally bound (ignoring external pressure, magnetic, and mass flow across the core boundary terms). The virial parameter is defined as, 
\begin{equation}
    \alpha = \left| \frac{2\Omega_K}{\Omega_G} \right| = \frac{5\sigma^2 R_{eff}}{GM}
\end{equation}
where $R_{eff}$ is the core effective radius, $M$ is mass of the core, $\sigma_v$ is the velocity dispersion from the ammonia observations and $G$ is the gravitational constant. The effective radius is defined as, 
\begin{equation}
R_{eff} = \sqrt{A/\pi}
\end{equation}
where $A$ is the area as defined from the ammonia NH$_3$ (1,1) intensity maps as described by \citealt{2015ApJ...805..185S}. The mass is calculated within the appropriate core area using the \textit{Herschel} column density map (subtracting off background). See section\,\ref{sourceselect} for further discussion on source size and extraction from \textit{Herschel} maps. Cores with methanol abundances $< 1.0 \times 10^{-9}$ are the cores considered `gravitationally bound' by the $\alpha$ parameter (Figure \,\ref{abund_corenum}, right).
Cores that are less gravitationally bound have had less time to collapse and thus are considered less dynamically evolved. 
This result also agrees with the chemical evolution expected for methanol.
Recent studies suggest that many starless cores are actually confined by external pressure, not their own gravity \citep{2018arXiv180910223C}. A full virial anaylsis combined with radiative transfer models of methanol observations of the cores are beyond the scope of this current paper and will be discussed in detail in a subsequent paper.

Abundance measurements for each core versus volume density within our beam are plotted in Figure \,\ref{textvsnbeam}. Volume density is also a potential evolutionary indicator, although it is not the sole evolutionary parameter as cores can evolve at different rates \citep{2005ApJ...632..982S}. We preface that all of the abundance plot comparisons have low correlation coefficients ($|r| < 0.26$). However, in general there are higher N(CH$_3$OH)/N(H$_2$) values at lower volume densities and there is more scatter across volume densities at lower N(CH$_3$OH)/N(H$_2$) values. 
We have listed in Table\,\ref{Abundances} all abundance measurements for each core, including NH$_3$ measurements from \citealt{2015ApJ...805..185S}. Comparing the N(NH$_3$)/N(CH$_3$OH) ratio, a late time chemical tracer, to $n_{beam}$ the ratio scatters toward higher abundance with increasing volume density (Figure \,\ref{textvsnbeam}).

\begin{figure*}
\centering
\includegraphics[width=185mm]{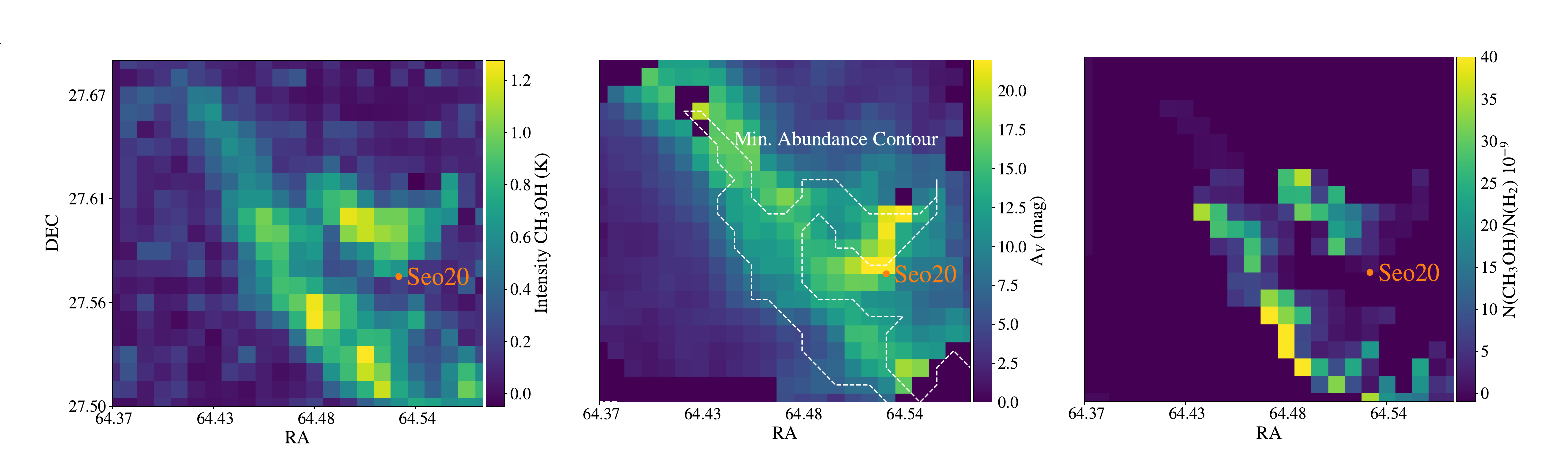}
\caption{ \label{abundmaps} Re-gridded CH$_3$OH brightness ($T_{mb}$), A$_\mathrm{V}$ and abundance maps, each convolved to 75$^{\prime\prime}$ resolution, for the B211 region of Taurus (left to right). The abundance map was created from points that were at least 5$\sigma$ detections in intensity.
}
\end{figure*}

\subsection{CH$_3$OH Spatial Distribution}\label{otfmapping}

We mapped seven 15$^{\prime}$ $\times$ 15$^{\prime}$ regions within the Taurus filament in the 96.7 GHz transitions of methanol, focusing on regions where our starless cores reside. These seven maps are named based on the Barnard region they lie in, i.e., B7, B10, B211, B213-1, B213-2, B216 and B218. In Figure \,\ref{extmap} we overlay the CH$_3$OH integrated intensity map contours on an extinction map from \citealt{2010ApJ...725.1327S} in steps of 0.2 K km s$^{-1}$ ($\sim$2$\sigma$ detection). The uniformly generated and Gaussian smoothed extinction map, at similar resolution as our CH$_3$OH beam ($\sim$ 1 arcminute), was generated from near-infrared (NIR) photometry ($JHK_s$ bands) of point sources throughout the Taurus L1495 filaments. We detected methanol emission at A$_\mathrm{V}$ as low as $\sim$ 3 mag (noise at $\sigma \sim 0.5$mag). In Table\,\ref{avregions} we quote the lowest extinction and $H_2$ column density values where methanol is detected at our $\sim$2$\sigma$ level in each region.

For every region, except B10, the integrated intensity maps were made within a velocity range from 3.19 to 8.03 km s$^{-1}$ where each channel was spaced by 0.12 km s$^{-1}$. In the case of B10 the range was from 5.09 to 7.51 km s$^{-1}$ spaced by 0.06 km s$^{-1}$. By using this cut off we focused only on the single brightest methanol transition, CH$_3$OH A 2$_0$-1$_0$.Regions denoted less-evolved, i.e., B211 and B10 in particular, show significant extended methanol emission (Figure \,\ref{extmap} and  \,\ref{velcontours}) in addition to having higher methanol abundances from the single pointing observations (Figure \,\ref{abund_corenum}). Even though our OTF integrated intensity maps are at modest angular resolution, we see indications of chemical differentiation. This can be clearly seen for Seo9 in the B10 region, which is one of the densest of the cores; i.e., the $\sim$7 km s$^{-1}$ methanol velocity component only slightly overlaps the Seo9 peak position (see Figure \,\ref{velcontours}).

We created a methanol abundance map of the `less-evolved' (i.e., no signs of protostars) B211 region. Gaussian line profiles were fit using CLASS at each point in the map, which is convolved to a finer resolution of 75$^{\prime\prime}$ (compared to 81$^{\prime\prime}$ in Figures \,\ref{extmap} and \,\ref{velcontours}).
We chose positions in our grid with integrated intensities of the brightest (CH$_3$OH A$^+$ 2$_{0}$-1$_0$) line that lie above the 5$\sigma$ rms level and run these points through a RADEX grid which calculates column densities and abundances (compared to H$_2$ from the $Herschel$ maps) for CH$_3$OH. The B211 region was the only region with enough points above 5$\sigma$ to create a reliable, spatially-connected abundance map (see Figure \,\ref{AVscatter}).
In Figure \,\ref{abundmaps} we present three of our re-gridded maps of peak brightness ($T_{mb}$), A$_\mathrm{V}$, and abundance. The peaks in the abundance map do not correspond to where core Seo20 is located, i.e., we find higher abundances along the filament than for the starless core itself. 
Extended emission in the filaments is also brighter than what was detected toward the NH$_3$-peak core positions by an order of magnitude in most other regions. This anti-correlation between bright methanol and dust emission toward core Seo20 is a clear sign of depletion, even given our modest spatial resolution, seen on larger filament-size scales. 

Putting together the trends discussed, we conclude that chemical differentiaion of methanol due to depletion in the central regions of cores is occurring. As suggested by the chemical desorption models of \cite{2017ApJ...842...33V}, methanol should preferentially be found in a shell around the dense central regions, where visual extinctions are large enough to screen interstellar UV photons ($\geq$ 10 mag) and volume densities are around a few $\times 10^4$ cm$^{-3}$. In fact, higher resolution observations of methanol towards more chemically evolved dense cores (L1498, L1517B; L1544; \citealt{2006A&A...455..577T}; \citealt{2014A&A...569A..27B}) have already revealed such ring-like structures in methanol emission.

\textbf{\begin{deluxetable}{lcc}
\tablecaption{ \label{avregions}}
\tablewidth{0pt}
\tablehead{ \colhead{OTF Mapped Region}& \colhead{A$_\mathrm{V}$}& \colhead{N$_{H_2}$ }  \\ \colhead{ }& \colhead{(mag)}& \colhead{(10$^{21}$ cm$^{-2}$)} }
\startdata
\hline
B7   &  11   &  6 \\
B10    &  5  &  4  \\
B211 & 4 & 3 \\
B213$^*$ & 5 & 3 \\
B216 & 4 & 3 \\
B218 & 3 & 2 \\
\enddata
\tablecomments{The values reported are the A$_\mathrm{V}$ and N$_{H_2}$ for which CH$_3$OH emission is detected at the 2$\sigma$ level from OTF maps. $^*$Including both B213-1 and B213-2 maps.}
\end{deluxetable}}

\begin{figure}[tbh]
\centering
\includegraphics[width=75mm]{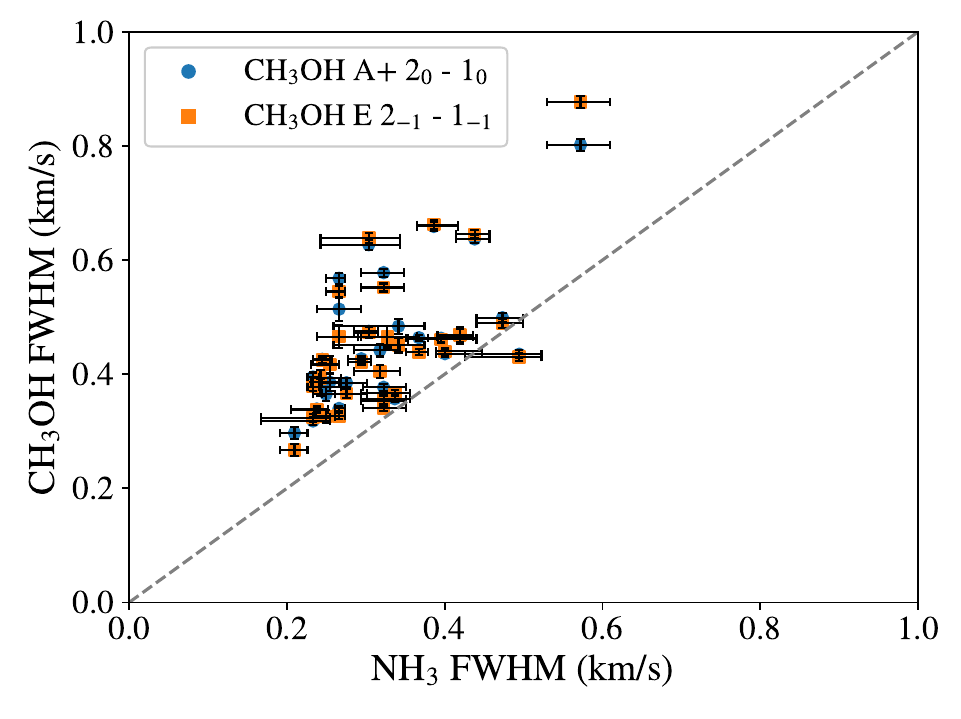}\\
\includegraphics[width=75mm]{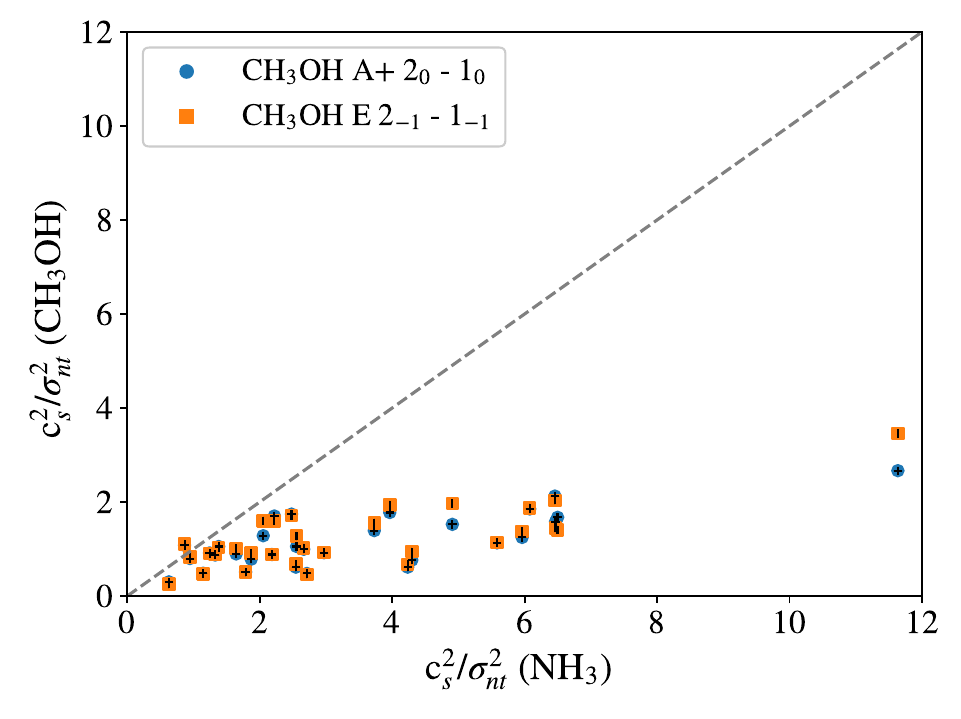} \\
\centering 
\includegraphics[width=75mm]{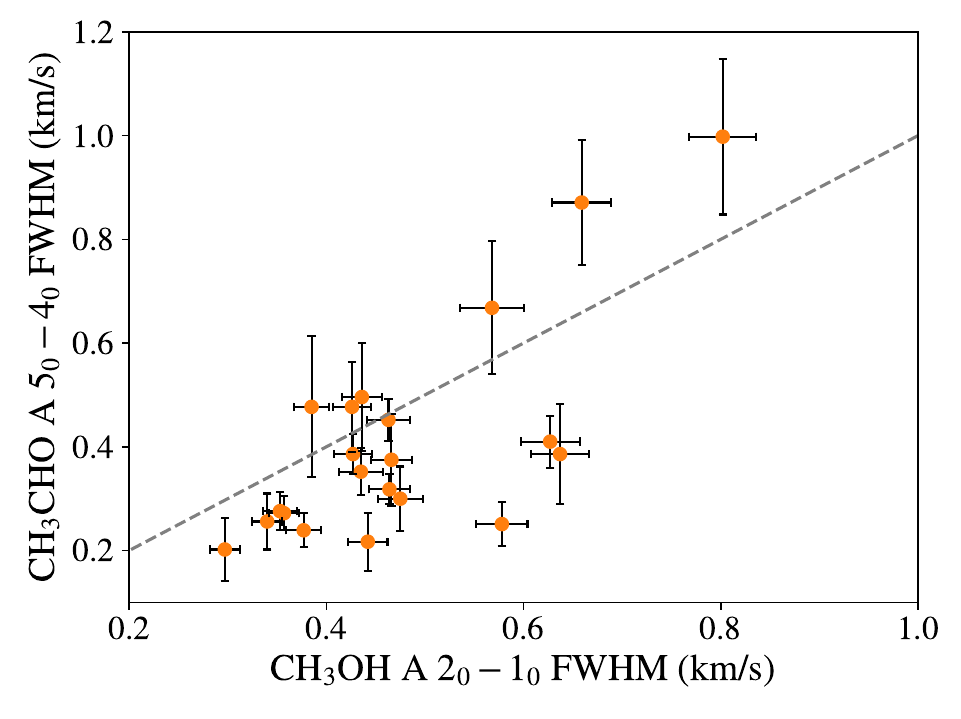}
\caption{\label{linestuff} (Top panel) Observed methanol linewidths  versus ammonia linewidths from \citealt{2015ApJ...805..185S}. (Middle panel) Plot of the ratio of thermal support to nonthermal support for methanol versus ammonia. (Bottom panel) We compare linewidth of methanol and acetaldehyde, finding a median linewidth ratio (CH$_3$CHO/CH$_3$OH) of 0.62. }
\end{figure}

\subsubsection{Multiple Velocity Peaks and Large Scale Motions} 

There are 3 cores ($\sim10\%$ of the sample) which are spatially nearby (Seo17, Seo20, and Seo21) and have shown clear evidence of multiple velocity components in our single pointing observations (see Figure \,\ref{taurus_met_spec}). Multiple velocity components have been seen in other molecular lines in this same region, i.e., C$^{18}$O(1-0) and N$_2$H$^+$(1-0) \citep{2013A&A...554A..55H}. They find two previously known velocity components (near 5.3 and 6.7 km s$^{-1}$), and we find a similar components at roughly the same v$_{LSR}$. Specifically, in core Seo20 each component peaks at 5.1 and 6.9 km s$^{-1}$, respectively. Multiple velocity components for methanol were seen clearly in not just the single pointings but in the OTF maps. In the B211 region a clear spatial separation of emission is found, i.e., in the $\sim$5 km s$^{-1}$ velocity channel we see methanol tracing the filament and in the $\sim$7 km s$^{-1}$ velocity channel methanol is centered around the starless cores Seo18/19, which were not included in our single pointing survey due to overlapping beams (Figure \,\ref{velcontours}). Observations in B10 also showed clear spatial separation of velocity components, telling us the methanol in these starless cores is not all coming from the same velocity structure, instead from multiple velocity channels at $\sim$5 km s$^{-1}$, $\sim$6 km s$^{-1}$ and $\sim$7 km s$^{-1}$ (Figure \,\ref{velcontours}). 

Perhaps it is not surprising that we observe multiple velocity components. In TMC-1 it is well known that two or more velocity components exist and that the line profile of CH$_3$OH is significantly broader than those of other molecules \citep{2015ApJ...802...74S}. Also, 17 dense cores in \cite{2018arXiv180205378T} were found to have multiple velocity peaks in $^{12}$CO, and in the case of Seo21 (which they map) the peaks occur at 5.40 and 7.49 km s$^{-1}$ which is close to what we observe in CH$_3$OH. The $\sim$5 km s$^{-1}$ velocity peak corresponds to that of the extended filament emission that \citealt{2013A&A...554A..55H} observed in C$^{18}$O. Thus, CH$_3$OH is tracing multiple parts of the cloud/filament along some line-of-sights. 

\begin{figure*}[tbh]
\centering
\includegraphics[scale=0.50]{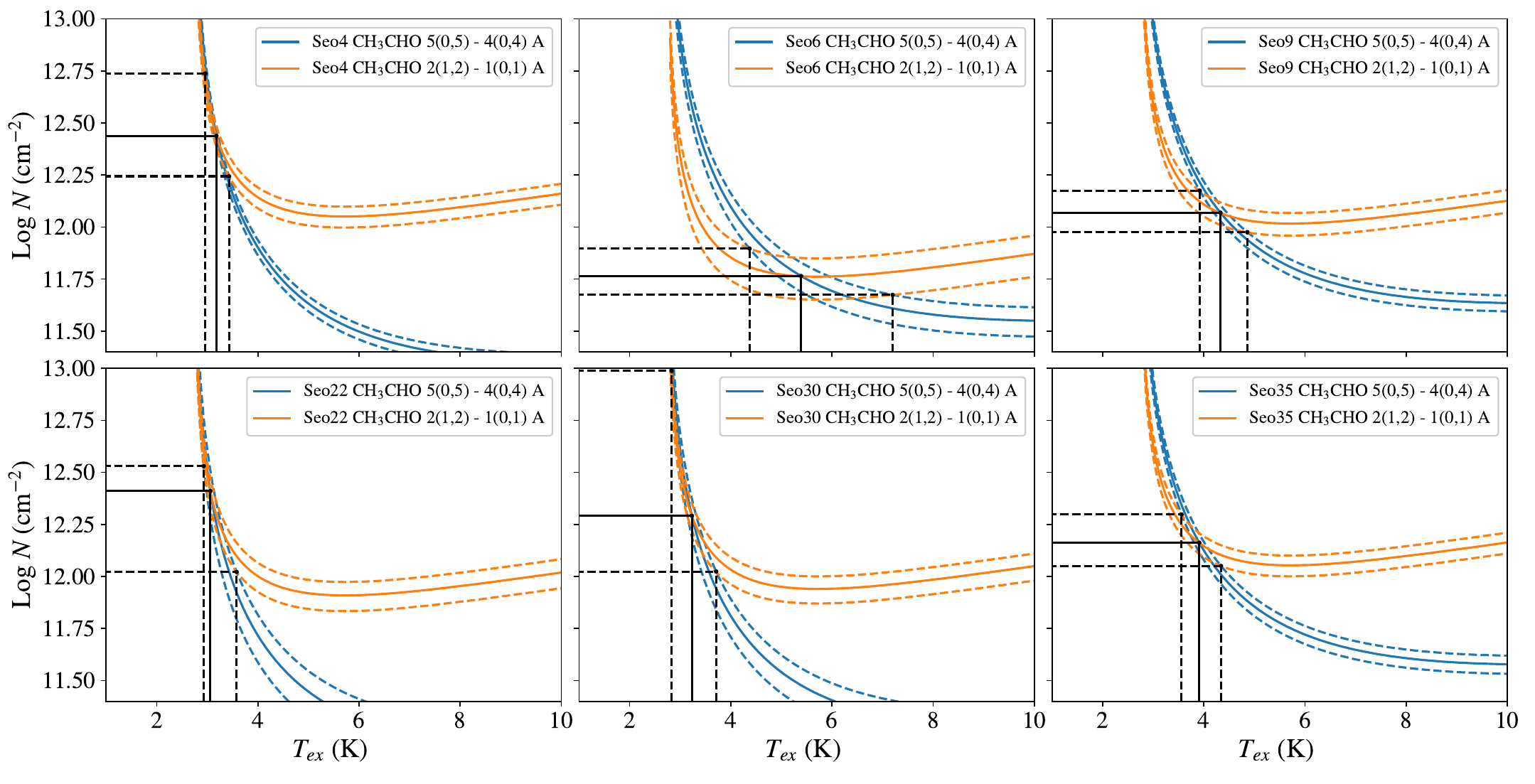}
\caption{\label{ch3chocTex} Log column density vs. excitation temperature calculated for two transitions of CH$_3$CHO using the CTEX method. Curves for each CH$_3$CHO transition were calculated given the observed integrated intensities and are plotted for all six of the cores for whch both transitions of CH$_3$CHO was detected. Intersecting values are recorded in Table\,\ref{ColDensities_acet}. }
\end{figure*}

In addition to the spatial separations in velocity, the CH$_3$OH linewidths have revealed a combination of unresolved bulk motions (gradients or flows) and supersonic turbulence. Using just the two brightest transition lines (A$^+$ 2$_{0}$-1$_0$ and E 2$_{-1}$-1$_{-1}$) for comparisons, we found that the linewidths of our CH$_3$OH lines were broader, at an average of $\sim$ 0.45 km s$^{-1}$ wide, than those of NH$_3$ observed by \citealt{2015ApJ...805..185S} (top panel of Figure \,\ref{linestuff}). The ratio of thermal support to non-thermal support was also smaller on average for CH$_3$OH than for NH$_3$, telling us that the methanol has a larger non-thermal contribution (middle panel of Figure \,\ref{linestuff}). Methanol emission is optically thin (section\,\ref{metcoldensec}), therefore optical depth is not the culprit for the wider linewidths. The nonthermal linewidth difference between NH$_3$ and CH$_3$OH could certainly arise from their difference
in sampling different densities of gas and therefore different large scale motions along the line-of-sight. 

In ($\sim 30\%$) of the cores there is evidence for non-Gaussian line asymmetries or `wings'; for example from its spectrum core Seo12 appears to have a red-shifted wing whereas Seo16 has a blue-shifted wing (Figure \,\ref{taurus_met_spec}). Regardless, for our line analysis (Gaussian fitting) we were only concerned with comparing the central velocity component ($\sim$ 7 km s$^{-1}$) at the v$_{LSR}$ of the cores.
Line asymmetries most likely represent a mixture of large-scale motions from within the core as well as the surrounding material which we detected within our large (62.3$^{''}$) beam.

\subsection{CH$_3$CHO Column Densities} \label{acet}

We detected acetaldehyde in 22 out of the 31 cores, where 18 out of the 22 were observed with at least 4$\sigma$ confidence, at $rms$ values $\sim$ 4-6 mK (Fig\,\ref{taurus_acet_spec} and Table\,\ref{ACETAFits}). Starless cores Seo11 and Seo20 are unique in that we detected the A but not the E transition state. Additionally, for core Seo21 we detected the E transition but not the A transition. In all other cases because both the A and E line were in the bandpass, and since they have similar upper energy levels ($E_u/k= 13.935$ K for CH$_3$CHO E and $E_u/k= 13.838$ K for CH$_3$CHO A), we could confirm CH$_3$CHO with a single spectrum. The cores with the highest main beam temperature were Seo32, Seo35, Seo9, Seo5 and Seo4 (from brightest to weakest). 

We found 6 out of the 22 cores detected in the $5_{(0,5)} - 4_{(0,4)}$ transition were also detectable in the 84 GHz CH$_3$CHO A $2_{(1,2)}-1_{(0,1)}$ transition, after integrating down to $rms$ values $\sim$ 3-4 mK (Table\,\ref{ch3choc_spectra} and Figure \,\ref{ch3choc_spectra}). Since CH$_3$CHO has no calculated collisional rate coefficients, RADEX calculations are not possible. We used the CTEX method which required at least two transitions with different $E_u/k$ values to simultaneously constrain $T_{ex}$ and the column density, $N$ (see Equation 80 of \citealt{2015PASP..127..266M}). Both the $5_{(0,5)} - 4_{(0,4)}$ and $2_{(1,2)}-1_{(0,1)}$ transitions were used to calculate $N$ and $T_{ex}$ for the 6 cores we detected both transitions in. The ranges for the six cores are $N$ = 1.2 -- 5.8 $\times$ 10$^{12}$ cm$^{-2}$ and $T_{ex}$ = 3.1 -- 5.4 K (Figure \,\ref{ch3chocTex} and Table\,\ref{ColDensities_acet}).

We extrapolated our results from CTEX in order to estimate the column densities for the remaining 16 cores where only the CH$_3$CHO $5_{(0,5)} - 4_{(0,4)}$ transition was detected. We used the median excitation temperature of the 6 cores, $T_{ex}$ = 3.57 K, and calculated the column density at that temperature. The total range of column densities for all 22 cores is 0.65 -- 5.8 $\times$ 10$^{12}$ cm$^{-2}$ (Table \,\ref{ColDensities_acet}).

\subsubsection{CH$_3$CHO Abundance Trends}

Previous studies have searched for acetaldehyde in only a handful of dense cores, including detections toward L183, TMC-1, CB17, L1689B, and L1544 (\citealt{1999ApJ...518..699T}, \citealt{2012A&A...541L..12B}, \citealt{2016ApJ...830L...6J}).
\citealt{2014ApJ...795L...2V} calculate a CH$_3$CHO column density of of 5.0 $\times10^{11}$ for L1544, however they assume $T_{ex}$ of 17 K. At a more realistic $T_{ex}$ of 5 K, \citealt{2016ApJ...830L...6J} report a column density of 1.2 $\times$ 10$^{12}$ cm$^{-2}$ at the center of L1544.
The CH$_3$CHO column density at $T_{ex}$ of 5 K for another very dense core, L1689B, for the E state 5 -- 4 transitions were found to be 9.12$\pm$0.92$\times$ 10$^{12}$ cm$^{-2}$, and for the A state 5 -- 4 transition 8.26$\pm$0.84$\times$ 10$^{12}$ cm$^{-2}$ \citep{2012A&A...541L..12B}. Our results suggest our our cores lie in between these core estimates. 

Acetaldehyde abundances compared to methanol, [CH$_3$CHO]/[CH$_3$OH], range from 0.02 -- 0.26 for the Taurus cores presented here. We note that many of the non-detections of CH$_3$CHO come from the B213 region, the same region where we detected multiple velocity components and lower  abundances of CH$_3$OH ($\lessapprox 1.5 \times 10^{-9}$).  Additionally, B213 is one of the most evolved regions with multiple embedded protostars. In the right panel of Figure \,\ref{abund_corenum} we show that cores with non-detections of CH$_3$CHO all have methanol abundances $< 1.5 \times 10^{-9}$ and that as abundances (wrt H$_2$) drop the larger the virial parameter (i.e., the more evolved the core). As cores evolve the abundance of both CH$_3$OH and CH$_3$CHO declines, suggesting that the formation processes for these two molecules are linked.

\subsection{CH$_3$CHO Linewidths}

In general the linewidths of the CH$_3$CHO A transition are narrower than those of methanol, with the average FWHM $\sim$ 0.23 km s$^{-1}$ wide (bottom panel of Fig\,\ref{linestuff}). The narrower linewidths indicate that the acetaldehyde is likely not tracing the full extent of A$_\mathrm{V}$ that is being traced by the methanol within the beam. The measured v$_{LSR}$ of CH$_3$CHO and CH$_3$OH are (on average) within $\sim$0.1 km s$^{-1}$ of each other. \citealt{2018ApJ...854..116S} also find narrower line widths for CH$_3$CHO vs. CH$_3$OH in TMC-1. Unfortunately, the acetaldehyde emission is weak which made it unfeasible to map the emission in a reasonable amount of time.

\section{Discussion}\label{sec:discuss}

The main conclusion of this paper is that methanol and acetaldehyde are easily observable in the gas phase toward a large sample of starless and prestellar cores, with a range of densities and ages, in the Taurus Molecular Cloud. Both CH$_3$OH and CH$_3$CHO are prevalent (100\% and 70\% detection rates respectively) with high gas phase abundances ($\sim$ $10^{-10}$ to 10$^{-9}$ wrt H$_2$). Given typical phase lifetimes of prestellar cores with densities $\sim 10^5$ cm$^{-3}$ of a few $\times 10^5$ years (see Figure 7 of \citealt{2014prpl.conf...27A}), then, at a minimum COM formation predates the formation of a first hydrostatic core by many hundreds of thousands of years.
Thus, with subsequent COM depletion in the central regions during the evolution of prestellar cores into first hydrostatic cores, our results suggest that protoplanetary disks will be seeded with COMs that have formed during the prestellar phase. Here we discuss the link between methanol and acetaldehyde, addressing how they might chemically co-evolve.

From our OTF maps we detect CH$_3$OH down to A$_\mathrm{V}$ of $\sim$ 3 mag, roughly where CO ice begins to form. The formation of CO ice begins in the gas-phase, during the so-called catastrophic CO freezeout stage, when it accretes onto layers of water ice that has already formed, resulting in a CO-rich apolar ice coating (\citealt{1991ApJ...381..181T}, \citealt{2006A&A...453L..47P}, \citealt{2011ApJ...740..109O}). From both astrochemical modeling and observations, CO freezeout in cold cores has been shown to occur at densities similar to starless core densities, i.e., a few 10$^{5}$ cm$^{-3}$ (\citealt{2005A&A...435..177J}, \citealt{2013A&A...560A..41L}). Methanol formation is believed to follow this freezeout process since CO freezout is a pre-requisite for CH$_3$OH ice formation without energetic radiation \citep{2009A&A...508..275C}. According to chemical desorption models, radicals are then desorbed off the ice and dust grains which react in the gas-phase to form more complex oraganics, like CH$_3$CHO (\citealt{2013ApJ...769...34V}, \citealt{2014ApJ...795L...2V}). Observations from \citealt{2014ApJ...795L...2V} support this idea, finding more complex organics, in addition to precursor methanol, are likely coming from an outer shell ($\sim$8000 AU for L1544) in a region where the ices are desorbed through non-thermal processes.

There are few possible gas-phase reactions which will form CH$_3$CHO in cold prestellar core environments. In one scenario, CH$_3$CHO is formed in the gas-phase by oxidation of the ethyl radical (C$_2$H$_5$ + O $\rightarrow$ CH$_3$CHO + H), as described by \citealt{2004AdSpR..33...23C}. This reaction, however, is not likely in cold cores due to the negligible reactive desorption probability of more complex radicals, i.e., C$_2$H$_5$ is predicted to have a very low reactive desorption probability \citep{2016A&A...585A..24M}. The most likely gas-phase reaction occurs between the methylidyne radical, CH, and methanol, CH$_3$OH, to form CH$_3$CHO, along with a hydrogen atom \citep{2000PCCP....2.2549J}.  This reaction requires methanol to have already been chemically desorbed into the gas phase.
The chemical link between CH$_3$OH and CH$_3$CHO is supported by the models of \citealt{2017ApJ...842...33V}, which show strong similarity in radial abundance profiles of both species (see their Figure 8).

From our results, there is no significant evolutionary correlation between the abundance of CH$_3$CHO with respect to H$_2$ or CH$_3$CHO with respect CH$_3$OH with volume density (Figure \,\ref{textvsnbeam}). Furthermore, abundance trends in the range of scatter observed for CH$_3$OH are not observed for CH$_3$CHO; there is similar scatter in CH$_3$CHO abundance and abundance ratios across the average volume densities probed. One reason for this may be that theoretical models of chemical desorption predict that there should be an enhancement of more complex organics, such as acetaldehyde, at both early and late times of the core's chemical evolution (Figure 7 in \citealt{2017ApJ...842...33V}). Obtaining central densities for the cores may reduce this scatter, since our beam-averaged volume density measurements only probe the global properties with a limited range (factor of 4 in density).

\begin{figure}[tbh]
\includegraphics[scale=0.5]{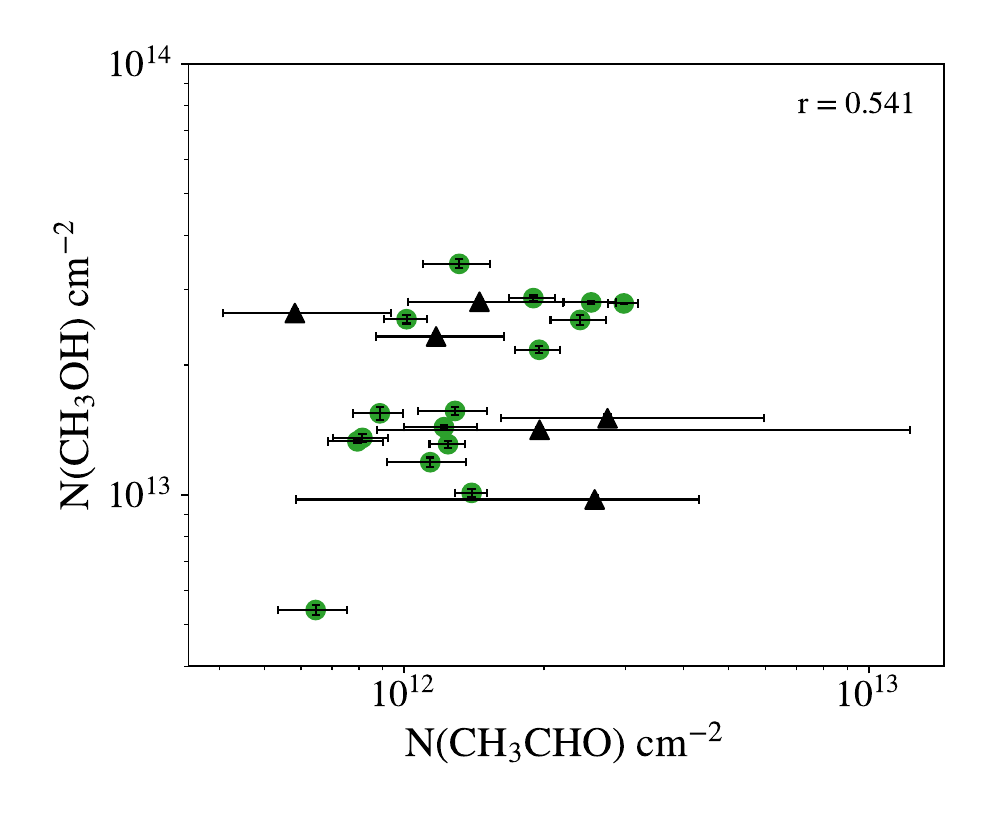}
\caption{\label{columndenboth} Column density of CH$_3$OH vs. CH$_3$CHO for the 22 cores for which CH$_3$CHO was detected. The black triangles symbolize the cores we detected both CH$_3$CHO transitions in, as in Figure \,\ref{abund_corenum}. The Spearman rank correlation coefficient is presented in the upper right corner.}
\end{figure}

We find cores with lower methanol abundances are less likely to be detected in CH$_3$CHO. All but two of the CH$_3$CHO non-detections are below the median methanol abundance (Figure\,\ref{abund_corenum}). The same trend with virial parameter for both CH$_3$OH and CH$_3$CHO is also found, that cores with smaller virial parameters (more evolved) have lower abundances (Figure\,\ref{abund_corenum}). When we plot the column density of CH$_3$OH vs. CH$_3$CHO (Figure \,\ref{columndenboth}) we find a weak but positive correlation ($r = 0.54$). These trends suggests that the CH + CH$_3$OH reaction is important for the gas-phase production of CH$_3$CHO in starless and prestellar cores. Still, with significant scatter in these trends, there may be other factors (i.e., beam filling fraction) affecting the relative abundances of CH$_3$OH and CH$_3$CHO that should to be addressed in future high resolution studies. Since we are still limited by our single-pointed observations, we cannot say whether strong chemical differentiation is occurring within our beam, although it seems highly likely. Obtaining higher spatial resolution maps of both species are needed to test against calculated radial profiles, spatial morphology's, and spatial scales at the core level.

\section{SUMMARY}\label{sec:conclude}

We found a prevalence of the organic molecules methanol (100\% detection rate) and acetaldehyde (70\% detection rate) toward a sample of 31 NH$_3$-identified starless and prestellar cores within the L1495-B218 filament in the Taurus Molecular Cloud. Our systematic survey shows that COMs, specifically methanol and acetaldehyde, that are important in prebiotic chemistry are forming early and often in the starless and prestellar stages at least hundreds of thousands of years prior to the formation of protostars and planets. We have calculated the column density, excitation temperature and abundance of CH$_3$OH and CH$_3$CHO for each core, comparing to physical properties, and we present maps of the distribution of methanol.

In all 31 cores we detected methanol, with total column densities ranging from 0.42 - 3.4 $\times$ 10$^{13}$ cm$^{-2}$ and excitation temperatures ranging from 6.79 - 8.66 K (from brightest transition). Additionally, in 22 out of 31 cores acetaldehyde was detected with column densities ranging from 0.65 - 5.81 $\times$ 10$^{12}$ cm$^{-2}$, with a median excitation temperature of 3.57 K. The total abundance of methanol spans from 0.53 -- 3.36 $\times 10^{-9}$ while the abundance of detected acetaldehyde spans 0.6 -- 3.9 $\times$ 10$^{-10}$ in the cores. Large scale motions are evident from asymmetric CH$_3$OH line profiles towards some cores. Multiple velocity components were seen in both the pointed observations as well as the OTF mapping of methanol that match well with the previously detected velocity coherent filament traced by C$^{18}$O (1-0). We find gas-phase methanol is an early time tracer, and was detected down to A$_\mathrm{V}$ as low as $\sim$ 3 mag. Analysis of the methanol observations are consistent with depletion in denser cores. There is evidence of a weak positive correlation between the abundances of methanol and acetaldehyde, however the chemical connection between these two molecules in prestellar cores has yet to be observationally supported by higher spatial resolution maps.

\section{Acknowledgements}

We would like to thank Youngmin Seo for his frequent assistance and the ARO 12m telescope operators (Michael Begam, Kevin Bays, Robert Thompson, and Clayton Kyle) who made these observations possible. We also thank our anonymous referee for their constructive comments. Samantha Scibelli is supported by National Science Foundation Graduate Research Fellowship (NSF GRF) Grant DGE-1143953.  Yancy Shirley and Samantha Scibelli were also supported in part by NSF Grant AST-1410190.

\software{\href{http://dx.doi.org/10.1051/0004-6361/201322068}{Astropy} (\citealt{2013A&A...558A..33A}, \citealt{2018AJ....156..123A}), \href{DOI:10.1109/MCSE.2011.37}{NumPy} \citep{2011CSE....13b..22V}, \href{http://www.scipy.org}{SciPy} \citep{2001OSST}, 
          \href{http://dx.doi.org/10.1109/MCSE.2007.55}{Matplotlib} \citep{2007CSE.....9...90H}, \href{https://ds9.si.edu}{Ds9} \citep{2003ASPC..295..489J}, \href{https://www.iram.fr/IRAMFR/GILDAS/}{GILDAS CLASS} (\citealt{2005sf2a.conf..721P}, \citealt{2013ascl.soft05010G}), \href{https://home.strw.leidenuniv.nl/~moldata/radex.html}{RADEX} \citep{2007A&A...468..627V}}

\begin{deluxetable}{ccccccc}
\tabletypesize{\tiny}
\tablecaption{Physical Parameters\label{physparams}}
\tablewidth{-5pt}
\tablehead{ \colhead{Core Number}&  $\alpha$ (J2000.0) &  $\delta$ (J2000.0) & \colhead{$n_{beam}$ (cm$^{-3}$)} &\colhead{$T_{dust}$ (K)} &\colhead{$T_{kin}$ (K)} }
\startdata
\hline
2& 4:18:33.1 & +28:27:11.0 &1.53E+05 &11.28 &9.66 $\SPSB{+0.09}{-0.08}$\\
4& 4:18:45.9 & +28:23:30.0 &1.57E+05 &13.06 &8.49 $\SPSB{+0.99}{-0.99}$\\
5& 4:18:04.2 & +28:22:51.6 &1.23E+05 &12.10 &9.65 $\SPSB{+1.52}{-1.34}$\\
6& 4:17:52.8 & +28:12:25.7 &1.18E+05 &11.46 &9.47 $\SPSB{+0.78}{-0.5}$\\
7& 4:18:02.6 & +28:10:27.8 &8.15E+04 &11.76 &9.48 $\SPSB{+0.43}{-0.55}$\\
8& 4:18:03.7 & +28:07:16.9 &9.50E+04 &11.42 &9.26 $\SPSB{+0.16}{-0.21}$\\
9& 4:18:07.0 & +28:05:13.0 &1.25E+05 &11.10 &8.61 $\SPSB{+0.3}{-0.24}$\\
10& 4:17:37.6 & +28:12:01.8 &6.84E+04 &12.39 &9.63 $\SPSB{+2.12}{-2.11}$\\
11& 4:17:51.2 & +28:14:25.0 &6.05E+04 &12.42 &10 $\SPSB{+0.39}{-0.54}$\\
12& 4:17:41.7 & +28:08:45.7 &1.31E+05 &11.17 &9.29 $\SPSB{+0.94}{-0.38}$\\
13& 4:17:42.2 & +28:07:29.4 &9.27E+04 &11.62 &10.9 $\SPSB{+1.06}{-1.41}$\\
14& 4:17:42.9 & +28:06:00.3 &9.72E+04 &11.65 &9.68 $\SPSB{+0.59}{-0.37}$\\
15& 4:17:41.0 & +28:03:49.9 &8.51E+04 &11.97 &8.84 $\SPSB{+0.39}{-0.5}$\\
16& 4:17:36.1 & +28:02:56.7 &9.66E+04 &11.81 &9.66 $\SPSB{+0.27}{-0.3}$\\
17& 4:17:50.3 & +27:55:52.4 &1.05E+05 &11.34 &9.31 $\SPSB{+0.59}{-0.54}$\\
20& 4:18:07.8 & +27:33:53.0 &9.16E+04 &11.59 &9.5 $\SPSB{+1.22}{-1.29}$\\
21& 4:19:23.3 & +27:14:46.0 &1.15E+05 &11.86 &9.17 $\SPSB{+1.36}{-1.3}$\\
22& 4:19:37.1 & +27:15:17.7 &1.19E+05 &11.35 &9.74 $\SPSB{+0.58}{-0.75}$\\
24& 4:19:51.4 & +27:11:26.3 &1.17E+05 &10.88 &9.08 $\SPSB{+0.15}{-0.16}$\\
26& 4:20:09.6 & +27:09:44.3 &5.86E+04 &11.95 &9.75 $\SPSB{+1.31}{-1.22}$\\
27& 4:20:14.7 & +27:07:38.8 &6.00E+04 &11.93 &10.3 $\SPSB{+1.01}{-1.09}$\\
28& 4:20:12.2 & +27:06:02.3 &5.73E+04 &11.86 &11.3 $\SPSB{+0.99}{-1.31}$\\
29& 4:20:15.4 & +27:04:23.2 &5.01E+04 &11.95 &10 $\SPSB{+0.67}{-0.97}$\\
30& 4:21:02.5 & +27:02:30.4 &9.08E+04 &11.42 &9.92 $\SPSB{+1.08}{-1.18}$\\
31& 4:20:51.6 & +27:01:53.6 &1.04E+05 &11.34 &9.55 $\SPSB{+0.49}{-0.52}$\\
32& 4:20:54.0 & +27:03:13.0 &1.28E+05 &11.09 &8.95 $\SPSB{+1.96}{-1.32}$\\
33& 4:21:21.6 & +26:59:30.6 &1.35E+05 &10.60 &9.54 $\SPSB{+0.93}{-0.58}$\\
35& 4:24:20.5 & +26:36:02.1 &8.97E+04 &11.34 &9.22$\SPSB{+0.8}{-1.11}$\\
36& 4:24:25.2 & +26:37:15.7 &7.24E+04 &11.69 &9.81 $\SPSB{+1.62}{-0.92}$\\
37& 4:27:47.4 & +26:17:57.8  &1.29E+05 &10.85 &9.83 $\SPSB{+0.5}{-0.62}$\\
39& 4:28:09.2 & +26:20:27.7 &1.71E+05 &10.56 &9.55 $\SPSB{+0.68}{-0.56}$\\
\enddata
\tablecomments{Physical parameters $Herschel$ column density and temperature maps, i.e., average volume density ($n_{beam}$) and average dust temperature ($T_{dust}$), as well as the gas kinetic temperature from ammonia maps taken from Table 2 of \citealt{2015ApJ...805..185S} ($T_{kin}$). }
\end{deluxetable}

\begin{deluxetable}{cccccccccccccc} 
\tabletypesize{\tiny}
\tablecaption{Methanol Gaussian Fit Results\label{MethFits}}
\tablewidth{-1pt}
\tabcolsep=0.08cm
\tablehead{ \colhead{}& \colhead{CH$_3$OH E 2$_0$ - 1$_0$}& \colhead{} &\colhead{} &\colhead{} & \colhead{CH$_3$OH A 2$_0$ - 1$_0$} &\colhead{}& \colhead{} &\colhead{}&  \colhead{CH$_3$OH E 2$_{-1}$ - 1$_{-1}$} &\colhead{}& \colhead{} &\colhead{} &\colhead{}\\
\colhead{Core} & \colhead{Area} & \colhead{Vel } & \colhead{FWHM} & \colhead{T$_{mb}$} & \colhead{Area} & \colhead{Vel } & \colhead{FWHM} & \colhead{T$_{mb}$} & \colhead{Area} & \colhead{Vel } & \colhead{FWHM} & \colhead{T$_{mb}$} & \colhead{$rms$}\\ 
\colhead{} & \colhead{ (K-km $s^{-1}$)} & \colhead{(km $s^{-1}$)} & \colhead{(km $s^{-1}$)} & \colhead{(K)} & \colhead{ (K-km $s^{-1}$)} & \colhead{(km $s^{-1}$)} & \colhead{(km $s^{-1}$)} & \colhead{(K)}& \colhead{ (K-km $s^{-1}$)} & \colhead{(km $s^{-1}$)} & \colhead{(km $s^{-1}$)} & \colhead{(K)} & \colhead{(K)}}
\startdata
2.0 &*0.026(0.005) &7.211(0.063) &0.607(0.184) &0.0402 &0.322(0.012) &7.248(0.008) &0.426(0.019) &0.71 &0.245(0.003) &7.2355(0.003) &0.425(0.006) &0.542 &0.0146\\
4.0 &0.0292(0.003) &7.191(0.013) &0.277(0.032) &0.0992 &0.325(0.013) &7.224(0.007) &0.377(0.018) &0.81 &0.238(0.003) &7.2165(0.002) &0.341(0.006) &0.656 &0.0138\\
5.0 &0.0408(0.005) &6.308(0.037) &0.703(0.1) &0.0545 &0.624(0.023) &6.367(0.015) &0.802(0.034) &0.731 &0.484(0.005) &6.3565(0.004) &0.877(0.011) &0.518 &0.013\\
6.0 &0.0306(0.004) &6.709(0.018) &0.318(0.046) &0.0905 &0.566(0.021) &6.723(0.008) &0.436(0.02) &1.22 &0.424(0.004) &6.7155(0.002) &0.44(0.005) &0.905 &0.0153\\
7.0 &0.018(0.003) &6.693(0.018) &0.253(0.043) &0.0666 &0.231(0.009) &6.67(0.006) &0.318(0.015) &0.684 &0.175(0.003) &6.6565(0.003) &0.324(0.007) &0.507 &0.0129\\
8.0 &0.0272(0.005) &6.725(0.026) &0.361(0.101) &0.0708 &0.299(0.011) &6.705(0.006) &0.34(0.015) &0.826 &0.224(0.003) &6.6905(0.002) &0.327(0.005) &0.643 &0.013\\
9.0 &0.0426(0.004) &6.903(0.016) &0.385(0.036) &0.104 &0.483(0.018) &6.866(0.008) &0.427(0.019) &1.06 &0.366(0.003) &6.8605(0.002) &0.421(0.004) &0.818 &0.0136\\
10.0 &0.0362(0.004) &5.481(0.017) &0.301(0.039) &0.113 &0.466(0.019) &5.525(0.008) &0.435(0.022) &1.01 &0.352(0.005) &5.5195(0.003) &0.43(0.007) &0.769 &0.016\\
11.0 &0.0472(0.004) &6.617(0.018) &0.417(0.042) &0.106 &0.563(0.022) &6.549(0.011) &0.578(0.026) &0.915 &0.416(0.004) &6.5385(0.003) &0.552(0.007) &0.707 &0.0139\\
12.0 &0.048(0.004) &5.919(0.017) &0.416(0.035) &0.108 &0.621(0.025) &5.933(0.012) &0.627(0.03) &0.93 &0.475(0.005) &5.9225(0.003) &0.639(0.009) &0.698 &0.0152\\
13.0 &0.049(0.005) &5.967(0.032) &0.684(0.072) &0.0673 &0.6(0.022) &6(0.012) &0.659(0.03) &0.855 &0.436(0.004) &5.9955(0.003) &0.662(0.008) &0.619 &0.013\\
14.0 &0.073(0.006) &5.998(0.033) &0.827(0.08) &0.0829 &0.755(0.028) &6.067(0.012) &0.637(0.029) &1.11 &0.558(0.005) &6.0535(0.003) &0.645(0.007) &0.814 &0.0147\\
15.0 &*0.0162(0.004) &6.724(0.031) &0.275(0.097) &0.0553 &0.328(0.013) &6.656(0.01) &0.484(0.025) &0.637 &0.233(0.005) &6.6545(0.005) &0.451(0.013) &0.485 &0.0149\\
16.0 &-- &-- &-- & -- &0.36(0.015) &6.592(0.012) &0.568(0.032) &0.595 &0.267(0.004) &6.5885(0.004) &0.545(0.01) &0.46 &0.0116\\
17.0 &*0.0279(0.005) &7.025(0.058) &0.569(0.111) &0.0461 &0.2(0.008) &6.898(0.007) &0.365(0.018) &0.514 &0.137(0.004) &6.8985(0.005) &0.326(0.012) &0.396 &0.017\\
20.0 &-- &-- &-- & -- &0.357(0.014) &6.876(0.009) &0.466(0.021) &0.72 &0.253(0.006) &6.8785(0.005) &0.469(0.013) &0.507 &0.0219\\
21.0 &*0.0245(0.004) &6.844(0.063) &0.649(0.112) &0.0354 &0.293(0.014) &6.685(0.011) &0.466(0.027) &0.59 &0.229(0.008) &6.6845(0.008) &0.465(0.019) &0.463 &0.0291\\
22.0 &-- &-- &-- & -- &0.23(0.009) &6.732(0.009) &0.442(0.02) &0.488 &0.167(0.004) &6.7365(0.005) &0.405(0.011) &0.388 &0.0143\\
24.0 &-- &-- &-- & -- &0.181(0.007) &6.554(0.007) &0.395(0.02) &0.43 &0.13(0.002) &6.5375(0.003) &0.378(0.008) &0.324 &0.0101\\
26.0 &*0.00588(0.001) &6.606(0.023) &0.213(0.059) &0.0259 &0.114(0.005) &6.688(0.011) &0.514(0.031) &0.208 &0.0794(0.002) &6.6445(0.006) &0.466(0.02) &0.16 &0.00706\\
27.0 &-- &-- &-- & -- &0.0957(0.004) &6.591(0.008) &0.377(0.021) &0.239 &0.0753(0.002) &6.5875(0.005) &0.393(0.015) &0.18 &0.00737\\
28.0 &*0.00807(0.002) &6.653(0.036) &0.299(0.09) &0.0253 &0.109(0.005) &6.655(0.008) &0.386(0.022) &0.264 &0.0887(0.003) &6.6465(0.006) &0.417(0.015) &0.2 &0.00882\\
29.0 &-- &-- &-- & -- &0.125(0.006) &6.54(0.006) &0.297(0.015) &0.396 &0.0895(0.003) &6.5315(0.005) &0.267(0.01) &0.314 &0.0141\\
30.0 &*0.0262(0.004) &6.786(0.041) &0.583(0.108) &0.0422 &0.324(0.013) &6.791(0.009) &0.475(0.023) &0.64 &0.246(0.004) &6.7775(0.003) &0.473(0.009) &0.488 &0.0124\\
31.0 &0.0238(0.003) &6.491(0.029) &0.4(0.055) &0.056 &0.376(0.014) &6.679(0.009) &0.499(0.022) &0.707 &0.27(0.004) &6.6695(0.003) &0.489(0.008) &0.519 &0.0134\\
32.0 &0.0501(0.004) &7.029(0.016) &0.417(0.034) &0.113 &0.565(0.023) &7.012(0.009) &0.463(0.022) &1.15 &0.453(0.005) &6.9995(0.002) &0.461(0.006) &0.924 &0.0169\\
33.0 &*0.027(0.005) &6.618(0.031) &0.411(0.098) &0.0617 &0.273(0.011) &6.567(0.007) &0.385(0.018) &0.667 &0.198(0.004) &6.5575(0.003) &0.366(0.008) &0.509 &0.0157\\
35.0 &0.0454(0.005) &6.679(0.021) &0.403(0.046) &0.106 &0.583(0.023) &6.682(0.009) &0.464(0.021) &1.18 &0.446(0.004) &6.6675(0.002) &0.439(0.005) &0.954 &0.0166\\
36.0 &0.0225(0.004) &6.552(0.022) &0.25(0.043) &0.0844 &0.293(0.012) &6.508(0.007) &0.353(0.017) &0.78 &0.22(0.004) &6.4955(0.003) &0.356(0.008) &0.581 &0.0156\\
37.0 &*0.0147(0.003) &6.973(0.045) &0.522(0.106) &0.0265 &0.234(0.009) &6.904(0.006) &0.357(0.015) &0.616 &0.174(0.003) &6.9005(0.003) &0.367(0.006) &0.445 &0.00992\\
39.0 &0.0343(0.004) &6.764(0.026) &0.473(0.069) &0.0682 &0.305(0.012) &6.759(0.006) &0.339(0.016) &0.846 &0.233(0.003) &6.7545(0.002) &0.337(0.005) &0.648 &0.0135\\
\enddata
\tablecomments{Gaussian fits for the three methanol lines observed. Errors reported in parentheses next to the number. *Upper limits ($<4\sigma_{rms}$)}.
\end{deluxetable}

\begin{deluxetable*}{cccccccccc}
\tablecaption{Acetaldehyde $5_{(0,5)} - 4_{(0,4)}$ Gaussian Fit Results\label{ACETAFits} }
\tablewidth{0pt}
\tablehead{ \colhead{}& \colhead{CH$_3$CHO A $5_{(0,5)} - 4_{(0,4)}$ }& \colhead{} &\colhead{} &\colhead{} &\colhead{CH$_3$CHO E $5_{(0,5)} - 4_{(0,4)}$} &\colhead{}& \colhead{} &\colhead{} &\colhead{} \\
\colhead{Core} & \colhead{Area} & \colhead{Vel } & \colhead{FWHM} & \colhead{T$_{mb}$} & \colhead{Area} & \colhead{Vel } & \colhead{FWHM} & \colhead{T$_{mb}$} &\colhead{$rms$}\\ 
\colhead{} & \colhead{ (K-km $s^{-1}$)} & \colhead{(km $s^{-1}$)} & \colhead{(km $s^{-1}$)} & \colhead{(K)} & \colhead{ (K-km $s^{-1}$)} & \colhead{(km $s^{-1}$)} & \colhead{(km $s^{-1}$)} & \colhead{(K)}&\colhead{(K)}}
\startdata
2 &0.0112(0.002) &7.379(0.043) &0.477(0.087) &0.022 &0.0125(0.002) &7.241(0.031) &0.415(0.051) &0.0284 &0.00662\\
4 &0.0119(0.001) &7.257(0.015) &0.239(0.033) &0.0469 &0.00963(0.001) &7.172(0.014) &0.208(0.036) &0.0435 &0.00713\\
5 &0.0232(0.003) &6.451(0.05) &0.998(0.15) &0.0218 &0.0274(0.003) &6.174(0.053) &1.08(0.145) &0.0238 &0.00573\\
6 &0.0187(0.003) &6.782(0.033) &0.496(0.104) &0.0354 &0.0123(0.002) &6.669(0.024) &0.326(0.048) &0.0355 &0.00816\\
7 & --  & -- & -- &-- &-- & -- &-- &-- &0.00744\\
8 &0.00747(0.001) &6.732(0.022) &0.256(0.054) &0.0274 &*0.0139(0.002) &6.844(0.097) &1.04(0.186) &*0.0126 &0.00595\\
9 &0.0227(0.002) &6.858(0.019) &0.386(0.039) &0.0554 &0.0179(0.003) &6.774(0.042) &0.494(0.109) &0.0341 &0.00826\\
10 &0.0179(0.002) &5.526(0.022) &0.352(0.045) &0.0479 &0.0169(0.002) &5.513(0.032) &0.454(0.065) &0.0351 &0.0083\\
11 &0.00929(0.001) &6.598(0.018) &0.251(0.042) &0.0348 &-- & -- &-- &-- &0.0066\\
12 &0.0174(0.002) &5.982(0.025) &0.41(0.05) &0.0399 &0.0302(0.003) &5.808(0.04) &0.799(0.088) &0.0355 &0.00789\\
13 &*0.022(0.003) &6.035(0.051) &0.871(0.121) &0.0237 &*0.0299(0.005) &6.122(0.132) &1.8(0.398) &0.0156 &0.00716\\
14 &*0.0121(0.002) &6.016(0.034) &0.386(0.097) &0.0293 &*0.0223(0.003) &6.009(0.055) &0.894(0.126) &0.0234 &0.00795\\
15 & --  & -- & -- &-- &-- & -- &-- &-- &0.00937\\
16 &*0.0118(0.002) &6.463(0.067) &0.668(0.128) &0.0166 &*0.0138(0.002) &6.392(0.06) &0.685(0.156) &0.0189 &0.00608\\
17 & --  & -- & -- &-- &-- & -- &-- &-- &0.00608\\
20 &*0.00814(0.001) &6.966(0.034) &0.375(0.089) &0.0204 &--& -- &-- &-- &0.00558\\
21 & -- & -- &-- &-- & 0.00728(0.001) &6.347(0.022) &0.235(0.044) &0.0291 &0.00671\\
22 &0.00779(0.002) &6.41(0.023) &0.217(0.056) &0.0337 &0.00957(0.002) &6.381(0.025) &0.277(0.059) &0.0324 &0.00773\\
24 & --  & -- & -- &-- &-- & -- &-- &-- &0.00769\\
26 & --  & -- & -- &-- &-- & -- &-- &-- &0.00769\\
27 & --  & -- & -- &-- &-- & -- &-- &-- &0.00727\\
28 & --  & -- & -- &-- &-- & -- &-- &-- &0.00702\\
29 &0.00592(0.001) &4.71(0.023) &0.202(0.06) &0.0275 &0.00502(0.001) &6.115(0.019) &0.165(0.04) &0.0285 &0.00728\\
30 &0.00966(0.002) &6.892(0.029) &0.3(0.062) &0.0303 &0.00704(0.002) &6.763(0.025) &0.218(0.049) &0.0304 &0.00821\\
31 & --  & -- & -- &-- &-- & -- &-- &-- &0.00776\\
32 &0.0273(0.002) &7.053(0.017) &0.452(0.041) &0.0567 &0.0276(0.003) &7.007(0.02) &0.457(0.054) &0.0567 &0.0077\\
33 &0.0105(0.002) &5.833(0.046) &0.477(0.136) &0.0206 &0.0114(0.001) &5.829(0.022) &0.342(0.04) &0.0314 &0.00594\\
35 &0.02(0.002) &6.684(0.014) &0.318(0.029) &0.059 &0.0219(0.002) &6.63(0.016) &0.336(0.033) &0.0613 &0.00787\\
36 &0.0114(0.001) &6.245(0.016) &0.276(0.036) &0.0389 &0.00697(0.001) &6.19(0.019) &0.216(0.046) &0.0304 &0.00622\\
37 &0.0128(0.001) &6.721(0.015) &0.272(0.033) &0.0442 &0.0185(0.002) &6.637(0.02) &0.429(0.038) &0.0405 &0.00609\\
39 & --  & -- & -- &-- &-- & -- &-- &-- &0.0069\\
\enddata
\tablecomments{Gaussian fits for the two acetaldehyde lines observed in the 95.9GHz range. Errors reported in parentheses next to the number. *Upper limits ($\lessapprox 4\sigma_{rms}$). A total of 18 out of the 31 cores had one or more of the $5_{(0,5)} - 4_{(0,4)}$ lines detected with significance. }
\end{deluxetable*}

\begin{deluxetable*}{ccccccccc}
\tabletypesize{\tiny}
\tablecaption{Acetaldehyde (2$_0$-1$_0$) Gaussian Fit Results\label{lineprops_acet}}
\tablewidth{0pt}
\tabcolsep=0.2cm
\tablehead{
\colhead{Core} & \colhead{Area} & \colhead{Vel} & \colhead{FWHM} & \colhead{T$_{mb}$} & \colhead{$rms$}  
\\
\colhead{ } & \colhead{ (K-km $s^{-1}$)} & \colhead{(km $s^{-1}$)} & \colhead{(km $s^{-1}$)} & \colhead{(K)} & \colhead{(K)} 
}
\startdata
4 &   8.68E-03(0.001)  &  7.143 (0.029) & 0.418 (0.054) & 1.950E-02 & 4.28E-03  \\
6 &  4.46E-03(0.001) & 6.647 (0.024)&  0.230 (0.056) & 1.82E-02 & 4.13E-03  \\
9 & 8.03E-03 (0.001) & 6.766 (0.037)  & 0.414 (0.080)  & 1.82E-02 & 5.13E-03  \\
22 & 6.27E-03 (0.001) &  6.54 (0.038)  & 0.403 (0.085)  & 1.46E-02  & 4.254E-03 \\
30 &  6.73E-03 (0.001) & 6.595 (0.043) & 0.447(0.102) & 1.42E-02 & 4.270E-03   \\
35 & 8.74E-03 (0.001) & 6.550 (0.046) & 0.493 (0.093)  & 1.67E-02& 4.277E-03 \\
\enddata
\tablecomments{See Fig\,\ref{ch3choc_spectra}.}
\end{deluxetable*} 

\begin{longrotatetable}
\begin{deluxetable*}{lllllllllll}
\tabletypesize{\tiny}
\tablecaption{Methanol Column Densities (cm$^{-2}$), Excitation temperatures (K) and Optical Depths $\tau$ \label{N_tex_tau}}
\tabcolsep=0.1cm
\tablehead{ \colhead{Core}&  \colhead{CH$_3$OH A\,\ ${2_0 - 1_0}$ } & \colhead{} & \colhead{} & \colhead{CH$_3$OH E\,\ ${2_{-1} - 1_{-1} }$} & \colhead{} & \colhead{} & \colhead{CH$_3$OH E\,\ ${2_{0} - 1_{0} }$} & \colhead{} & \colhead{} & \colhead{$Sum_{A+E}$($Std$)}  \\ 
\colhead{}&  \colhead{$N$ } & \colhead{$T_{ex}$} & \colhead{$\tau$} & \colhead{$N$} & \colhead{$T_{ex}$} & \colhead{$\tau$} & \colhead{$N$} & \colhead{$T_{ex}$} & \colhead{$\tau$} & \colhead{$N_{tot}$}  
\\
\colhead{ }&  \colhead{10$^{13}$ cm$^{-2}$} & \colhead{K} & \colhead{} &\colhead{10$^{13}$ cm$^{-2}$}&  \colhead{K} & \colhead{} &  \colhead{10$^{13}$ cm$^{-2}$} & \colhead{K} & \colhead{} &\colhead{10$^{13}$ cm$^{-2}$}  }
\startdata
2 &0.735$\SPSB{+0.03}{-0.03}$&8.33$\SPSB{+0.415}{-0.41}$ &0.1358$\SPSB{+0.0055}{-0.0055}$ &0.7$\SPSB{+0.015}{-0.01}$ &8.328$\SPSB{+0.001}{-<0.001}$ &0.1101$\SPSB{+0.0023}{-0.0016}$ &*0.71$\SPSB{+0.18}{-0.19}$ &8.329$\SPSB{+2.02}{-2.03}$ &0.01251$\SPSB{+0.00321}{-0.00337}$  &1.4(0.035)\\
4 &0.79$\SPSB{+<0.001}{-0.025}$ &7.393$\SPSB{+0.324}{-0.327}$ &0.1941$\SPSB{+<0.001}{-0.0061}$ &0.715$\SPSB{+0.005}{-0.03}$ &7.388$\SPSB{+0.001}{-0.002}$ &0.1666$\SPSB{+0.0011}{-0.0069}$ &0.85$\SPSB{+<0.001}{-0.01}$ &7.397$\SPSB{+1.57}{-1.57}$ &0.03653$\SPSB{+<0.001}{-<0.001}$ &1.5(0.075)\\
5 &1.425$\SPSB{+0.01}{-0.045}$ &8.079$\SPSB{+0.47}{-0.472}$ &0.1452$\SPSB{+0.0011}{-0.0046}$ &1.37$\SPSB{+0.005}{-0.05}$ &8.077$\SPSB{+<0.001}{-0.002}$ &0.1108$\SPSB{+<0.001}{-0.004}$ &1.22$\SPSB{+0.035}{-0.02}$ &8.072$\SPSB{+2.1}{-2.1}$ &0.01895$\SPSB{+<0.001}{-<0.001}$  &2.8(0.055)\\
6 &1.375$\SPSB{+0.015}{-0.02}$ &7.926$\SPSB{+0.482}{-0.478}$ &0.2637$\SPSB{+0.0028}{-0.0037}$ &1.265$\SPSB{+0.01}{-0.01}$ &7.918$\SPSB{+<0.001}{-0.001}$ &0.2098$\SPSB{+0.0017}{-0.0016}$ &0.935$\SPSB{+0.05}{-0.07}$ &7.896$\SPSB{+2.04}{-2.05}$ &0.03293$\SPSB{+0.0018}{-0.00251}$ &2.6(0.11)\\
7 &0.535$\SPSB{+0.015}{-0.005}$ &7.369$\SPSB{+0.548}{-0.544}$ &0.1539$\SPSB{+0.0042}{-0.0014}$ &0.5$\SPSB{+<0.001}{-0.005}$ &7.364$\SPSB{+<0.001}{-<0.001}$ &0.1273$\SPSB{+<0.001}{-0.0013}$ &0.64$\SPSB{+0.06}{-0.065}$ &7.384$\SPSB{+2.06}{-2.07}$ &0.02944$\SPSB{+0.00279}{-0.00303}$  &1.0(0.035)\\
8 &0.705$\SPSB{+0.025}{-0.02}$ &7.457$\SPSB{+0.495}{-0.488}$ &0.1878$\SPSB{+0.0065}{-0.0053}$ &0.65$\SPSB{+0.005}{-<0.001}$ &7.45$\SPSB{+0.001}{-<0.001}$ &0.1602$\SPSB{+0.0012}{-<0.001}$ &0.925$\SPSB{+0.2}{-0.22}$ &7.483$\SPSB{+2.02}{-2.03}$ &0.02969$\SPSB{+0.00653}{-0.00714}$&1.4(0.055)\\
9 &1.21$\SPSB{+0.025}{-0.03}$ &7.304$\SPSB{+0.39}{-0.385}$ &0.2653$\SPSB{+0.0054}{-0.0064}$ &1.12$\SPSB{+0.005}{-0.005}$ &7.298$\SPSB{+<0.001}{-0.001}$ &0.2176$\SPSB{+0.001}{-0.001}$ &1.34$\SPSB{+0.04}{-0.045}$ &7.313$\SPSB{+1.72}{-1.72}$ &0.04156$\SPSB{+0.00127}{-0.00143}$  &2.3(0.09)\\
10 &1.125$\SPSB{+0.04}{-0.185}$ &7.231$\SPSB{+0.598}{-0.632}$ &0.2406$\SPSB{+0.0083}{-0.0383}$ &1.045$\SPSB{+0.055}{-0.18}$ &7.221$\SPSB{+0.007}{-0.022}$ &0.207$\SPSB{+0.0106}{-0.0345}$ &1.365$\SPSB{+0.01}{-0.11}$ &7.26$\SPSB{+2.1}{-2.1}$ &0.05407$\SPSB{+<0.001}{-0.00444}$ &2.2(0.08)\\
11 &1.335$\SPSB{+0.025}{-<0.001}$ &7.234$\SPSB{+0.662}{-0.658}$ &0.2136$\SPSB{+0.0039}{-<0.001}$ &1.22$\SPSB{+0.015}{-0.035}$ &7.222$\SPSB{+0.002}{-0.004}$ &0.189$\SPSB{+0.0023}{-0.0053}$ &1.835$\SPSB{+0.07}{-0.05}$ &7.287$\SPSB{+2.19}{-2.19}$ &0.0523$\SPSB{+0.00203}{-0.00144}$ &2.6(0.12)\\
12 &1.47$\SPSB{+0.02}{-0.04}$ &7.873$\SPSB{+0.433}{-0.429}$ &0.1986$\SPSB{+0.0026}{-0.0053}$ &1.39$\SPSB{+0.005}{-<0.001}$ &7.87$\SPSB{+<0.001}{-<0.001}$ &0.1594$\SPSB{+<0.001}{-<0.001}$ &1.415$\SPSB{+0.015}{-0.025}$ &7.871$\SPSB{+1.94}{-1.94}$ &0.03844$\SPSB{+<0.001}{-<0.001}$ &2.9(0.08)\\
13 &1.335$\SPSB{+0.03}{-0.015}$ &8.664$\SPSB{+0.649}{-0.648}$ &0.1498$\SPSB{+0.0033}{-0.0016}$ &1.21$\SPSB{+<0.001}{-0.035}$ &8.657$\SPSB{+<0.001}{-0.002}$ &0.1194$\SPSB{+<0.001}{-0.0034}$ &1.55$\SPSB{+0.045}{-<0.001}$ &8.676$\SPSB{+2.71}{-2.72}$ &0.02349$\SPSB{+<0.001}{-<0.001}$  &2.5(0.12)\\
14 &1.8$\SPSB{+0.035}{-0.035}$ &7.827$\SPSB{+0.545}{-0.539}$ &0.2393$\SPSB{+0.0045}{-0.0046}$ &1.635$\SPSB{+0.01}{-0.015}$ &7.818$\SPSB{+0.001}{-0.001}$ &0.1905$\SPSB{+0.0012}{-0.0017}$ &2.385$\SPSB{+0.06}{-0.08}$ &7.859$\SPSB{+2.22}{-2.22}$ &0.03243$\SPSB{+<0.001}{-0.00111}$ &3.4(0.17)\\
15 &0.78$\SPSB{+0.02}{-0.005}$ &6.982$\SPSB{+0.465}{-0.462}$ &0.1589$\SPSB{+0.004}{-0.001}$ &0.68$\SPSB{+0.005}{-0.005}$ &6.972$\SPSB{+0.001}{-0.001}$ &0.134$\SPSB{+<0.001}{-<0.001}$ &*0.595$\SPSB{+0.155}{-0.18}$ &6.963$\SPSB{+1.79}{-1.81}$ &0.02612$\SPSB{+0.0069}{-0.00796}$ &1.5(0.1)\\
16 &0.815$\SPSB{+0.035}{-0.035}$ &7.754$\SPSB{+0.521}{-0.515}$ &0.1234$\SPSB{+0.0053}{-0.0052}$ &0.75$\SPSB{+0.01}{-0.005}$ &7.749$\SPSB{+0.001}{-<0.001}$ &0.105$\SPSB{+0.0014}{-<0.001}$ &--&--&--&1.6(0.065)\\
17 &0.455$\SPSB{+0.01}{-0.01}$ &7.599$\SPSB{+0.462}{-0.459}$ &0.1106$\SPSB{+0.0024}{-0.0024}$ &0.385$\SPSB{+0.01}{-0.005}$ &7.591$\SPSB{+0.001}{-<0.001}$ &0.09242$\SPSB{+0.00237}{-0.00119}$ &*0.91$\SPSB{+0.1}{-0.115}$ &7.648$\SPSB{+2.05}{-2.05}$ &0.01816$\SPSB{+0.00202}{-0.00232}$&0.84(0.07)\\
20 &0.825$\SPSB{+<0.001}{-0.035}$ &7.567$\SPSB{+0.522}{-0.526}$ &0.157$\SPSB{+<0.001}{-0.0065}$ &0.72$\SPSB{+0.005}{-0.03}$ &7.558$\SPSB{+0.001}{-0.002}$ &0.1216$\SPSB{+<0.001}{-0.005}$ &--&--&--&1.5(0.1)\\
21 &0.675$\SPSB{+0.005}{-0.015}$ &7.612$\SPSB{+0.435}{-0.436}$ &0.1284$\SPSB{+<0.001}{-0.0028}$ &0.655$\SPSB{+0.005}{-0.015}$ &7.61$\SPSB{+0.001}{-0.001}$ &0.1093$\SPSB{+<0.001}{-0.0024}$ &*0.775$\SPSB{+0.04}{-0.03}$ &7.619$\SPSB{+1.94}{-1.95}$ &0.01353$\SPSB{+<0.001}{-<0.001}$ &1.3(0.02)\\
22 &0.51$\SPSB{+0.015}{-0.005}$ &8.093$\SPSB{+0.471}{-0.469}$ &0.09429$\SPSB{+0.00274}{-<0.001}$ &0.465$\SPSB{+0.01}{-<0.001}$ &8.09$\SPSB{+<0.001}{-<0.001}$ &0.08152$\SPSB{+0.00173}{-<0.001}$ &--&--&--&0.97(0.045)\\
24 &0.41$\SPSB{+0.02}{-0.015}$ &7.549$\SPSB{+0.419}{-0.415}$ &0.09316$\SPSB{+0.00451}{-0.00338}$ &0.365$\SPSB{+0.005}{-0.005}$ &7.545$\SPSB{+0.001}{-<0.001}$ &0.07583$\SPSB{+0.00103}{-0.00104}$ &--&--&--&0.78(0.045)\\
26 &0.255$\SPSB{+<0.001}{-0.005}$ &6.9$\SPSB{+0.608}{-0.609}$ &0.0488$\SPSB{+<0.001}{-<0.001}$ &0.22$\SPSB{+<0.001}{-0.005}$ &6.895$\SPSB{+<0.001}{-0.001}$ &0.04309$\SPSB{+<0.001}{-<0.001}$ &*0.235$\SPSB{+0.04}{-0.04}$ &6.898$\SPSB{+1.95}{-1.95}$ &0.01326$\SPSB{+0.00226}{-0.00227}$ &0.47(0.035)\\
27 &0.21$\SPSB{+0.005}{-<0.001}$ &7.269$\SPSB{+0.67}{-0.669}$ &0.05116$\SPSB{+0.00121}{-<0.001}$ &0.205$\SPSB{+<0.001}{-0.005}$ &7.268$\SPSB{+<0.001}{-0.001}$ &0.04433$\SPSB{+<0.001}{-0.00107}$ &--&--&--&0.41(0.005)\\
28 &0.23$\SPSB{+0.005}{-<0.001}$ &7.832$\SPSB{+0.789}{-0.788}$ &0.04949$\SPSB{+0.00106}{-<0.001}$ &0.235$\SPSB{+0.005}{-0.005}$ &7.833$\SPSB{+0.001}{-0.001}$ &0.04334$\SPSB{+<0.001}{-<0.001}$ &*0.305$\SPSB{+0.065}{-0.07}$ &7.845$\SPSB{+2.55}{-2.56}$ &0.01121$\SPSB{+0.0024}{-0.00258}$ &0.47(0.005)\\
29 &0.285$\SPSB{+0.005}{-0.005}$ &6.785$\SPSB{+0.659}{-0.659}$ &0.09569$\SPSB{+0.00165}{-0.00165}$ &0.255$\SPSB{+<0.001}{-0.005}$ &6.776$\SPSB{+<0.001}{-0.002}$ &0.08947$\SPSB{+<0.001}{-0.00173}$ &--&--&--&0.54(0.03)\\
30 &0.73$\SPSB{+0.01}{-0.01}$ &7.856$\SPSB{+0.563}{-0.563}$ &0.1295$\SPSB{+0.0018}{-0.0018}$ &0.685$\SPSB{+<0.001}{-0.025}$ &7.852$\SPSB{+<0.001}{-0.002}$ &0.1088$\SPSB{+<0.001}{-0.0039}$ &*0.875$\SPSB{+0.075}{-0.075}$ &7.869$\SPSB{+2.27}{-2.27}$ &0.01661$\SPSB{+0.00144}{-0.00143}$  &1.4(0.045)\\
31 &0.86$\SPSB{+0.025}{-0.01}$ &7.791$\SPSB{+0.496}{-0.493}$ &0.1476$\SPSB{+0.0042}{-0.0017}$ &0.765$\SPSB{+0.005}{-0.005}$ &7.784$\SPSB{+<0.001}{-<0.001}$ &0.1181$\SPSB{+<0.001}{-<0.001}$ &0.765$\SPSB{+0.05}{-0.05}$ &7.784$\SPSB{+2.1}{-2.1}$ &0.0214$\SPSB{+0.00142}{-0.00142}$ &1.6(0.095)\\
32 &1.395$\SPSB{+0.035}{-0.095}$ &7.6$\SPSB{+0.412}{-0.423}$ &0.2673$\SPSB{+0.0065}{-0.0179}$ &1.385$\SPSB{+0.045}{-0.105}$ &7.6$\SPSB{+0.003}{-0.006}$ &0.2314$\SPSB{+0.0073}{-0.0172}$ &1.52$\SPSB{+0.09}{-0.115}$ &7.607$\SPSB{+1.84}{-1.84}$ &0.04239$\SPSB{+0.00257}{-0.0033}$ &2.8(0.01)\\
33 &0.625$\SPSB{+0.01}{-0.015}$ &8.096$\SPSB{+0.431}{-0.429}$ &0.1327$\SPSB{+0.0021}{-0.0031}$ &0.565$\SPSB{+0.005}{-0.005}$ &8.092$\SPSB{+<0.001}{-<0.001}$ &0.1086$\SPSB{+<0.001}{-0.001}$ &*0.78$\SPSB{+0.11}{-0.135}$ &8.107$\SPSB{+2.04}{-2.05}$ &0.02077$\SPSB{+0.00298}{-0.00364}$ &1.2(0.06)\\
35 &1.445$\SPSB{+<0.001}{-0.075}$ &7.393$\SPSB{+0.509}{-0.518}$ &0.2845$\SPSB{+<0.001}{-0.0143}$ &1.36$\SPSB{+0.03}{-0.1}$ &7.386$\SPSB{+0.003}{-0.009}$ &0.2529$\SPSB{+0.0054}{-0.0181}$ &1.57$\SPSB{+0.035}{-0.015}$ &7.404$\SPSB{+2.02}{-2.02}$ &0.0459$\SPSB{+0.00105}{-<0.001}$ &2.8(0.085)\\
36 &0.68$\SPSB{+0.01}{-0.01}$ &7.414$\SPSB{+0.605}{-0.609}$ &0.1738$\SPSB{+0.0025}{-0.0025}$ &0.63$\SPSB{+0.015}{-0.02}$ &7.407$\SPSB{+0.003}{-0.002}$ &0.1454$\SPSB{+0.0034}{-0.0046}$ &0.825$\SPSB{+0.04}{-0.07}$ &7.435$\SPSB{+2.16}{-2.17}$ &0.03837$\SPSB{+0.00189}{-0.0033}$&1.3(0.05)\\
37 &0.525$\SPSB{+0.015}{-0.01}$ &8.275$\SPSB{+0.466}{-0.463}$ &0.1166$\SPSB{+0.0033}{-0.0022}$ &0.485$\SPSB{+0.01}{-<0.001}$ &8.272$\SPSB{+0.001}{-<0.001}$ &0.09034$\SPSB{+0.00184}{-<0.001}$ &*0.43$\SPSB{+0.055}{-0.06}$ &8.268$\SPSB{+2.17}{-2.17}$ &0.008782$\SPSB{+0.00113}{-0.00123}$ &1.0(0.04)\\
39 &0.71$\SPSB{+0.02}{-0.02}$ &8.35$\SPSB{+0.387}{-0.383}$ &0.1643$\SPSB{+0.0046}{-0.0046}$ &0.68$\SPSB{+0.01}{-0.005}$ &8.348$\SPSB{+0.001}{-<0.001}$ &0.1332$\SPSB{+0.0019}{-0.001}$ &0.895$\SPSB{+0.06}{-0.065}$ &8.36$\SPSB{+1.91}{-1.92}$ &0.02044$\SPSB{+0.0014}{-0.00151}$ &1.4(0.03)\\
\enddata
\tablecomments{We have listed column densities $N$, excitation temperature $T_{ex}$ and optical depth $\tau$ calculated for the three methanol transitions from brightest to weakest here. *Cores for which we only had upper limits for ($<4\sigma_{rms}$). For six of the cores the weakest transition was not detected and we only report the brightest two transitions.}
\end{deluxetable*}
\end{longrotatetable}

\begin{deluxetable}{cccc}
\tabletypesize{\tiny}
\tablecaption{Acetaldehyde: $N$ and $T_{ex}$\label{ColDensities_acet}}
\tablewidth{0pt}
\tabcolsep=1.0cm
\tablehead{
\colhead{Core} & \colhead{$N$} & \colhead{$T_{ex}$ } 
\\
\colhead{ } &\colhead{10$^{12}$ cm$^{-2}$} & \colhead{K}}
\startdata
4 &  2.741$\SPSB{+3.223}{-1.123}$ & 3.18$\SPSB{+0.25}{-0.22}$ \\
6 & 0.582$\SPSB{+0.357}{-0.175}$ & 5.39$\SPSB{+1.81}{-1.02}$ \\
9 &  1.172$\SPSB{+0.468}{-0.303}$ & 4.33$\SPSB{+0.53}{-0.41}$ \\
22 & 2.572$\SPSB{+1.735}{-1.987}$ & 3.06$\SPSB{+0.52}{-0.13}$\\
30 & 1.957$\SPSB{+10.293}{-1.082}$ & 3.23$\SPSB{+0.48}{-0.40}$  \\
35 & 1.453$\SPSB{+0.752}{-0.432}$ &3.91$\SPSB{+0.35}{-0.43}$ \\
\hline 
\hline 
Range & 1.172-5.812 & 3.06-5.39\\
\hline 
2 &  1.219$\SPSB{+0.218}{-0.218}$ & 3.57\\
5 &  2.528$\SPSB{+0.327}{-0.327}$ & 3.57\\
8 &  0.814$\SPSB{+0.11}{-0.11}$ & 3.57\\
10&  1.953$\SPSB{+0.218}{-0.218}$ & 3.57\\
11&  1.012$\SPSB{+0.109}{-0.109}$ & 3.57\\
12&  1.898$\SPSB{+0.218}{-0.218}$ & 3.57\\
*13&  2.393$\SPSB{+0.327}{-0.327}$ & 3.57\\
*14&  1.314$\SPSB{+0.218}{-0.218}$ & 3.57\\
*16&  1.287$\SPSB{+0.218}{-0.218}$ & 3.57\\
*20&  0.887$\SPSB{+0.11}{-0.11}$ & 3.57\\
21&  0.793$\SPSB{+0.11}{-0.11}$ & 3.57\\
29&  0.645$\SPSB{+0.11}{-0.11}$ & 3.57\\
32&  2.970$\SPSB{+0.218}{-0.218}$ & 3.57\\
33&  1.138$\SPSB{+0.218}{-0.218}$ & 3.57\\
36&  1.243$\SPSB{+0.11}{-0.11}$ & 3.57\\
37& 1.398$\SPSB{+0.11}{-0.11}$ & 3.57\\
\hline 
\hline 
Range & 0.65-5.81 & --\\
\enddata
\tablecomments{Table split up into first the six cores where both transitions were detected (see Figure \,\ref{ch3chocTex}) and then the remaining 16 cores we extrapolate the column densities from the median $T_{ex}$ of the six cores. Note: the estimate for core 21 is from the E state, the rest are from the brighter A state line. We quote both the range of column densities for just the six cores as well as for all 22 cores. }
\end{deluxetable} 

\begin{deluxetable*}{cllllll}
\centering 
\tablecaption{Abundances\label{Abundances}}
\tablewidth{0pt}
\tabcolsep=0.4cm
\tablehead{
\colhead{Core} & \colhead{N(CH$_3$OH)/N(H$_2$)} & \colhead{N(CH$_3$CHO)/N(H$_2$)} & \colhead{N(CH$_3$CHO)/N(CH$_3$OH)} & \colhead{N(NH$_3$)/N(H$_2$)}
\\
\colhead{ } & \colhead{10$^{-9}$ } & \colhead{10$^{-9}$} & \colhead{} & \colhead{10$^{-9}$} }
\startdata
2 &0.745$\SPSB{+0.00826}{-0.0101}$&0.0633$\SPSB{+0.0103}{-0.0126}$&0.0849$\SPSB{+0.0138}{-0.0169}$&1.89(0.394)\\
4 &0.762$\SPSB{+0.0173}{-0.0211}$&0.139$\SPSB{+0.148}{-0.0632}$&0.182$\SPSB{+0.195}{-0.0831}$&0.306(0.1)\\
5 &1.8$\SPSB{+0.0161}{-0.0197}$&0.163$\SPSB{+0.0191}{-0.0234}$&0.0904$\SPSB{+0.0107}{-0.013}$&0.605(0.124)\\
6 &1.78$\SPSB{+0.0337}{-0.0412}$&0.0392$\SPSB{+0.0219}{-0.0131}$&0.022$\SPSB{+0.0123}{-0.00738}$&0.941(0.192)\\
7 &1.01$\SPSB{+0.0155}{-0.019}$& -- & -- &1.6(0.245)\\
8 &1.13$\SPSB{+0.0209}{-0.0256}$&0.068$\SPSB{+0.00828}{-0.0101}$&0.06$\SPSB{+0.00739}{-0.00904}$&1.95(0.32)\\
9 &1.48$\SPSB{+0.0259}{-0.0317}$&0.0743$\SPSB{+0.027}{-0.0213}$&0.0503$\SPSB{+0.0183}{-0.0145}$&1.82(0.442)\\
10 &2.52$\SPSB{+0.0422}{-0.0516}$&0.227$\SPSB{+0.023}{-0.0281}$&0.09$\SPSB{+0.00925}{-0.0113}$&1.05(0.35)\\
11 &3.36$\SPSB{+0.0687}{-0.0839}$&0.133$\SPSB{+0.013}{-0.0159}$&0.0396$\SPSB{+0.00396}{-0.00484}$& -- \\
12 &1.74$\SPSB{+0.0221}{-0.027}$&0.116$\SPSB{+0.0121}{-0.0147}$&0.0664$\SPSB{+0.00698}{-0.00853}$&1.74(0.238)\\
13 &2.18$\SPSB{+0.0487}{-0.0595}$&*0.205$\SPSB{+0.0255}{-0.0311}$&*0.094$\SPSB{+0.0119}{-0.0145}$&0.908(0.206)\\
14 &2.81$\SPSB{+0.0613}{-0.0749}$&*0.107$\SPSB{+0.0162}{-0.0198}$&*0.0383$\SPSB{+0.00583}{-0.00712}$&0.923(0.185)\\
15 &1.36$\SPSB{+0.0425}{-0.0519}$& -- & -- &1.03(0.198)\\
16 &1.29$\SPSB{+0.0243}{-0.0297}$&*0.106$\SPSB{+0.0163}{-0.0199}$&*0.0823$\SPSB{+0.0128}{-0.0156}$&1.28(0.215)\\
17 &0.636$\SPSB{+0.0241}{-0.0295}$& -- & -- &2.26(0.342)\\
20 &1.34$\SPSB{+0.0414}{-0.0506}$&*0.0769$\SPSB{+0.00859}{-0.0105}$&*0.0574$\SPSB{+0.00665}{-0.00813}$&0.699(0.247)\\
21 &0.916$\SPSB{+0.00626}{-0.00766}$&0.0547$\SPSB{+0.00682}{-0.00834}$&0.0597$\SPSB{+0.00746}{-0.00911}$&0.791(0.2)\\
22 &0.653$\SPSB{+0.0137}{-0.0167}$&0.172$\SPSB{+0.106}{-0.148}$&0.264$\SPSB{+0.162}{-0.227}$&2.27(0.429)\\
24 &0.525$\SPSB{+0.0139}{-0.0169}$& -- & -- &4.06(0.465)\\
26 &0.644$\SPSB{+0.0216}{-0.0264}$& -- & -- &1.06(0.313)\\
27 &0.55$\SPSB{+0.00301}{-0.00368}$& -- & -- &0.898(0.26)\\
28 &0.645$\SPSB{+0.00315}{-0.00386}$& -- & -- &1.38(0.353)\\
29 &0.857$\SPSB{+0.0216}{-0.0264}$&0.102$\SPSB{+0.0157}{-0.0192}$&0.119$\SPSB{+0.0186}{-0.0227}$&1.41(0.35)\\
30 &1.24$\SPSB{+0.0179}{-0.0219}$&0.171$\SPSB{+0.819}{-0.105}$&0.138$\SPSB{+0.661}{-0.085}$&1.77(0.328)\\
31 &1.24$\SPSB{+0.0329}{-0.0402}$& -- & -- &0.862(0.14)\\
32 &1.72$\SPSB{+0.00282}{-0.00344}$&0.184$\SPSB{+0.0123}{-0.015}$&0.107$\SPSB{+0.00713}{-0.00871}$&0.471(0.129)\\
33 &0.701$\SPSB{+0.0161}{-0.0196}$&0.0671$\SPSB{+0.0117}{-0.0143}$&0.0957$\SPSB{+0.0168}{-0.0205}$&3.8(0.53)\\
35 &2.48$\SPSB{+0.0342}{-0.0418}$&0.129$\SPSB{+0.0606}{-0.0425}$&0.0518$\SPSB{+0.0244}{-0.0171}$&0.736(0.233)\\
36 &1.44$\SPSB{+0.0249}{-0.0305}$&0.136$\SPSB{+0.0109}{-0.0133}$&0.0949$\SPSB{+0.00774}{-0.00946}$&0.822(0.308)\\
37 &0.622$\SPSB{+0.0112}{-0.0137}$&0.0861$\SPSB{+0.0061}{-0.00745}$&0.138$\SPSB{+0.0101}{-0.0124}$&3.27(0.446)\\
39 &0.644$\SPSB{+0.00632}{-0.00773}$& -- & -- &2.35(0.326)\\
\hline 
\enddata
\tablecomments{Methanol column density used to calculate abundances is total (A+E state) value reported in column 11 of Table\,\ref{N_tex_tau}. N(NH$_3$)/N(H$_2$) from abundance maps provided by \citealt{2015ApJ...805..185S} which we use to calculate what the median ratio would be within our methanol beam size (standard deviation in parentheses). *Derived from upper limit CH$_3$CHO measurements.}
\end{deluxetable*}

\end{document}